\documentclass[fleqn,10pt]{wlscirep}
\usepackage[utf8]{inputenc}
\usepackage[T1]{fontenc}
\usepackage{color, textcomp}
\usepackage{xr-hyper}
\usepackage[colorlinks=true, allcolors=blue]{hyperref}
\DeclareMathOperator*{\argmin}{arg\,min}

\makeatletter
\newcommand*{\addFileDependency}[1]{\typeout{(#1)}
\@addtofilelist{#1}
\IfFileExists{#1}{}{\typeout{No file #1.}}
}\makeatother
\newcommand*{\myexternaldocument}[1]{%
\externaldocument{#1}%
\addFileDependency{#1.tex}%
\addFileDependency{#1.aux}%
}
\myexternaldocument{SI}

\title{AI-guided inverse design and discovery of recyclable vitrimeric polymers}

\author[1]{Yiwen Zheng}
\author[2]{Prakash Thakolkaran}
\author[1]{Agni K. Biswal}
\author[3,4]{Jake A. Smith}
\author[3]{Ziheng Lu}
\author[3]{Shuxin Zheng}
\author[3,4*]{Bichlien H. Nguyen}
\author[2*]{Siddhant Kumar}
\author[1*]{Aniruddh Vashisth}
\affil[1]{Department of Mechanical Engineering, University of Washington, Seattle, WA, USA}
\affil[2]{Department of Materials Science and Engineering, Delft University of Technology, Delft, The Netherlands}
\affil[3]{Microsoft Research AI4Science}
\affil[4]{Paul G. Allen School of Computer Science and Engineering, University of Washington, Seattle, WA, USA}
\affil[*]{bnguy@microsoft.com, sid.kumar@tudelft.nl, vashisth@uw.edu}

\begin{abstract}
Vitrimer is a new, exciting class of sustainable polymers with the ability to heal due to their dynamic covalent adaptive network that can go through associative rearrangement reactions. However, a limited choice of constituent molecules restricts their property space, prohibiting full realization of their potential applications. To overcome this challenge, we couple molecular dynamics (MD) simulations and a novel graph variational autoencoder (VAE) machine learning model for inverse design of vitrimer chemistries with desired glass transition temperature ($T_\mathrm{g}$) and synthesize a novel vitrimer polymer. We build the first vitrimer dataset of one million chemistries and calculate $T_\mathrm{g}$ on 8,424 of them by high-throughput MD simulations calibrated by a Gaussian process model. The proposed novel VAE employs dual graph encoders and a latent dimension overlapping scheme which allows for individual representation of multi-component vitrimers. By constructing a continuous latent space containing necessary information of vitrimers, we demonstrate high accuracy and efficiency of our framework in discovering novel vitrimers with desirable $T_\mathrm{g}$ beyond the training regime. To validate the effectiveness of our framework in experiments, we generate novel vitrimer chemistries with a target $T_\mathrm{g}$ = 323 K. By incorporating chemical intuition, we synthesize a vitrimer with $T_\mathrm{g}$ of 311-317 K, and experimentally demonstrate healability and flowability. The proposed framework offers an exciting tool for polymer chemists to design and synthesize novel, sustainable vitrimer polymers for a facet of applications.
\end{abstract}

\begin{document}
\flushbottom
\maketitle
\makeabstract
\thispagestyle{empty}

\section*{Introduction}

\begin{figure}[t]
\centering
\includegraphics[width=\linewidth]{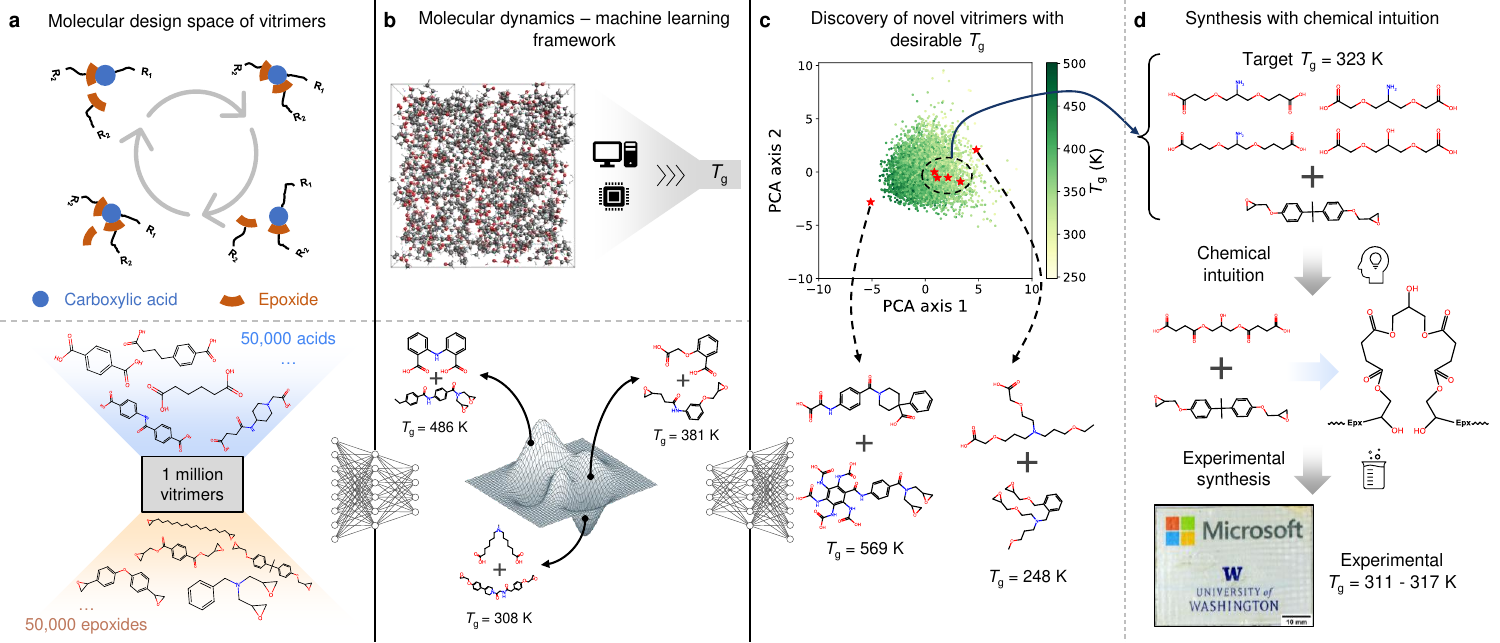}
\caption{\textbf{Schematic overview of this work.} \textbf{(a)} A transesterification vitrimer comprises a carboxylic acid and an epoxide. The reversible covalent bond between acid and epoxide allows them to detach from and attach to each other, thus healing the polymer. The design space for vitrimers is defined as all possible combinations of 50,000 carboxylic acids and 50,000 epoxides and a vitrimer dataset is built by sampling from the design space. \textbf{(b)} We use calibrated MD simulations to calculate $T_\mathrm{g}$ on a subset of vitrimers. The vitrimer dataset and $T_\mathrm{g}$ are inputs to the VAE model. \textbf{(c)} By optimizing latent vectors according to desirable $T_\mathrm{g}$, novel vitrimers with $T_\mathrm{g}$ = 569 K and 248 K are discovered. \textbf{(d)} Synthesis of novel vitrimer chemistry proposed by the framework for target $T_\mathrm{g}$ of 323 K (50 \textdegree C).}
\label{overview}
\end{figure}

Polymers are essential to a broad range of applications from cars and wind turbines to smartphones, medical devices and more; however their performance decreases over their life-cycle initiated by bond breaking at the molecular scale due to high stress, oxidation, or other factors. Mechanical damage due to the rupturing of covalent bonds in traditional thermosets and thermoplastics is irreversible, resulting in crack formation and finally failure \cite{young2011introduction}. In such circumstances, plastics end up in waste streams due to the inability to serve the desired purpose, which presents two key challenges for sustainability. First, failure in repairing mechanical damage means polymer parts and often entire assemblies must be replaced, resulting in high economic cost and further increasing the 430 million tons of plastic produced annually. Second, the inability of polymers to repair molecular damage poses a fundamental challenge preventing recycling of thermosets altogether and the degradation of thermoplastics such as PET water bottles into highly degraded secondary raw materials. 

Healable polymers, particularly a new class called vitrimers, offer a potential solution to the plastic waste problem. Combining durability with end-of-life recyclability, vitrimers have the potential to greatly reduce the amount of plastic production and waste \cite{krishnakumar2020vitrimers}. The defining molecular feature of vitrimers is associative dynamic covalent adaptive network (CAN) which allows the constituents of polymer chains to attach to and detach from each other while conserving crosslinking density under an external stimulus such as heat. This gives vitrimers the ability of self-healing without loss of viscosity \cite{montarnal2011silica} (Figure \ref{overview}a). This exchange of constituents is termed a rearrangement reaction, and polymer scientists have found multiple reaction chemistries on which to base vitrimers, including transesterification, disulfide exchange, and imine exchange \cite{jin2019malleable}.  However, available vitrimers have restricted thermo-mechanical properties due to limited commercially available monomers (i.e., building blocks) for synthesizing these polymers, which is a key impediment to widespread applications of vitrimers.

The structure-property relationships of polymers have been primarily investigated in a forward manner: given a set of polymers, one queries their properties by experiments and simulations \cite{valavala2005modeling, vashisth2018effect}. At the early stage, most of the novel polymers are discovered and synthesized based on chemical intuition in a trial-and-error fashion \cite{hoogenboom2003combinatorial}. As chemical synthesis of polymers is expensive and time consuming, virtual specimen fabrication and characterization of desired chemical structures using molecular dynamics (MD) simulations may be employed to reduce the cost of experimentation. MD is a simulation technique situated at the interface of quantum mechanics and classical mechanics and has been widely employed to assist the discovery process \cite{hansson2002molecular}. Virtual characterization using MD has helped in gaining insights about the effect of polymer molecular structures on mechanical properties \cite{vashisth2018accelerated}, glass transition temperature \cite{yu2001polymer} and self-healing \cite{kamble2022reversing}. However, scaled computational screens assisted by MD or other simulation methods remain costly, even with the development of high-performance computing \cite{kranenburg2009challenges}. As a result, the searchable design space is limited to the order of $10^3$ to $10^5$ compositions.

Advances in machine learning (ML) algorithms offer an opportunity to accelerate polymer discovery by learning from available data, revealing hidden patterns in material properties \cite{guo2021artificial} and reducing the need for costly experiments and simulations \cite{barnett2020designing}. Various ML methods have been employed to design organic molecules and polymers, including forward predictive models \cite{jorgensen2018machine, tao2021machine, tao2021benchmarking, yang2022machine}, generative adversarial networks \cite{kadurin2017drugan, sanchez2017optimizing, prykhodko2019novo}, variational autoencoders (VAEs) \cite{gomez2018automatic, jin2018junction, jin2020hierarchical, batra2020polymers, jiang2024property}, diffusion models \cite{xu2022geodiff, hoogeboom2022equivariant} and large language models \cite{jablonka2024leveraging}. The trained ML models can be further used for high-throughput screening or conditioned upon physical properties to achieve the inverse design of polymers from properties of interest, such as glass transition temperature ($T_\mathrm{g}$)\cite{tao2021machine, tao2021benchmarking}, thermal conductivity\cite{wu2019machine, zhu2020machine, yang2022machine}, bandgap\cite{batra2020polymers} and gas-separation properties \cite{yao2021inverse, yang2022machine}. The success in these ML models depends on the choice of suitable representations, which is challenging due to the discrete and undefined degrees of freedom of molecules and polymers. To date, researchers have employed strings \cite{weininger1988smiles, krenn2020self}, molecular fingerprints \cite{rogers2010extended} and graphs \cite{kearnes2016molecular} to represent molecules and monomers in ML models. In this work, we propose a graph VAE model employing dual graph encoders and overlapping latent dimensions \cite{lerique2020joint, zheng2023unifying} which enable representation of multi-component vitrimers and controlled design of selective components simultaneously.

Last few years have seen an increased contribution to structures and properties databases of polymers and molecules such as ZINC15\cite{sterling2015zinc}, ChemSpider\cite{pence2010chemspider} and PubChem\cite{kim2023pubchem}. However, a dataset of vitrimers to train such a deep generative model is lacking. Furthermore, part of the dataset needs to be associated with the property of interest to enable property-guided inverse design. Vitrimers are characterized by two key thermal properties: glass transition temperature ($T_\mathrm{g}$) and topology freezing temperature ($T_\mathrm{v}$). $T_\mathrm{g}$ describes the transition from glassy state to rubbery state while $T_\mathrm{v}$ describes the transition from viscoelastic solids to viscoelastic liquids. At service temperatures, vitrimers perform like traditional polymers, but when heated to $T_\mathrm{v}$, the chains gain mobility and carry out exchange reactions at the reactive sites. Traditionally, vitrimer polymers exhibit $T_\mathrm{v}$ > $T_\mathrm{g}$ \cite{kamble2022reversing, van2020vitrimers}; this makes their future application easier since $T_\mathrm{g}$ dictates design protocols, and healing in vitrimers happens at temperatures higher than $T_\mathrm{g}$. Therefore, in this work, we focus our efforts on designing vitrimers with targeted $T_\mathrm{g}$. We build the first vitrimer dataset derived from the online database ZINC15 \cite{sterling2015zinc} and calculate $T_\mathrm{g}$ by calibrated MD simulations on a subset of vitrimers (Figure \ref{overview}a,b).

Leveraging this vitrimer dataset, we build an integrated MD-ML framework for discovery of bifunctional transesterification vitrimers with desirable properties specifically targeted $T_\mathrm{g}$ for the scope of this work (Figure \ref{overview}b). Each vitrimer contains two reactive constituents (i.e., carboxylic acid and epoxide). Furthermore, the discrete nature of the molecules prohibits a smooth and continuous design space. For example, while molecules are interpretable to human, they are not interpretable to a numerical optimizer for design of vitrimers. To this end, we develop a VAE that receives as input a vitrimer represented by graphs and subgraphs of the constituents and produces a smooth and continuous latent space. In such a latent space, two similar vitrimers are located close to each other while an optimizer can traverse the space of all possible vitrimers. Our unique VAE framework offers both constituent-specific and joint latent spaces of the chemical constituents, i.e., continuous screening and optimization can be performed on just one or both of the constituents. This enables interpretability on the effects of optimizing over, e.g., acid only, epoxide only, or simultaneously acid and epoxide molecules. The efficacy of the framework is demonstrated by discovering novel vitrimers with $T_\mathrm{g}$ both within and well beyond the dataset. Specifically, while the $T_\mathrm{g}$ in the training data ranges from 250 K to 500 K, we discover vitrimers with $T_\mathrm{g}$ around 569 K and 248 K (Figure \ref{overview}c). To validate the framework predictions, we generate vitrimer chemistries with a $T_\mathrm{g}$ of 323 K (50 \textdegree C). We choose this $T_\mathrm{g}$ range to develop vitrimers comparable to commonly used polyamides that find applications in transportation, electronics, consumer goods, and packaging. We examine the top candidates suggested by framework and use chemical intuition to further ensure simpler synthesizability and thermodynamic stability (Figure \ref{overview}d). The novel, synthesized vitrimer is characterized and experimental $T_\mathrm{g}$ is in good agreement with inverse design target $T_\mathrm{g}$. This validates the proposed framework, which is sufficiently general to be applied to different types of vitrimers and their properties, as a tool for polymer chemists to discover and synthesize novel vitrimer chemistries with desirable properties.

\section*{Results}

\subsection*{Design space and data generation}

\begin{figure}[t]
\centering
\includegraphics[width=\linewidth]{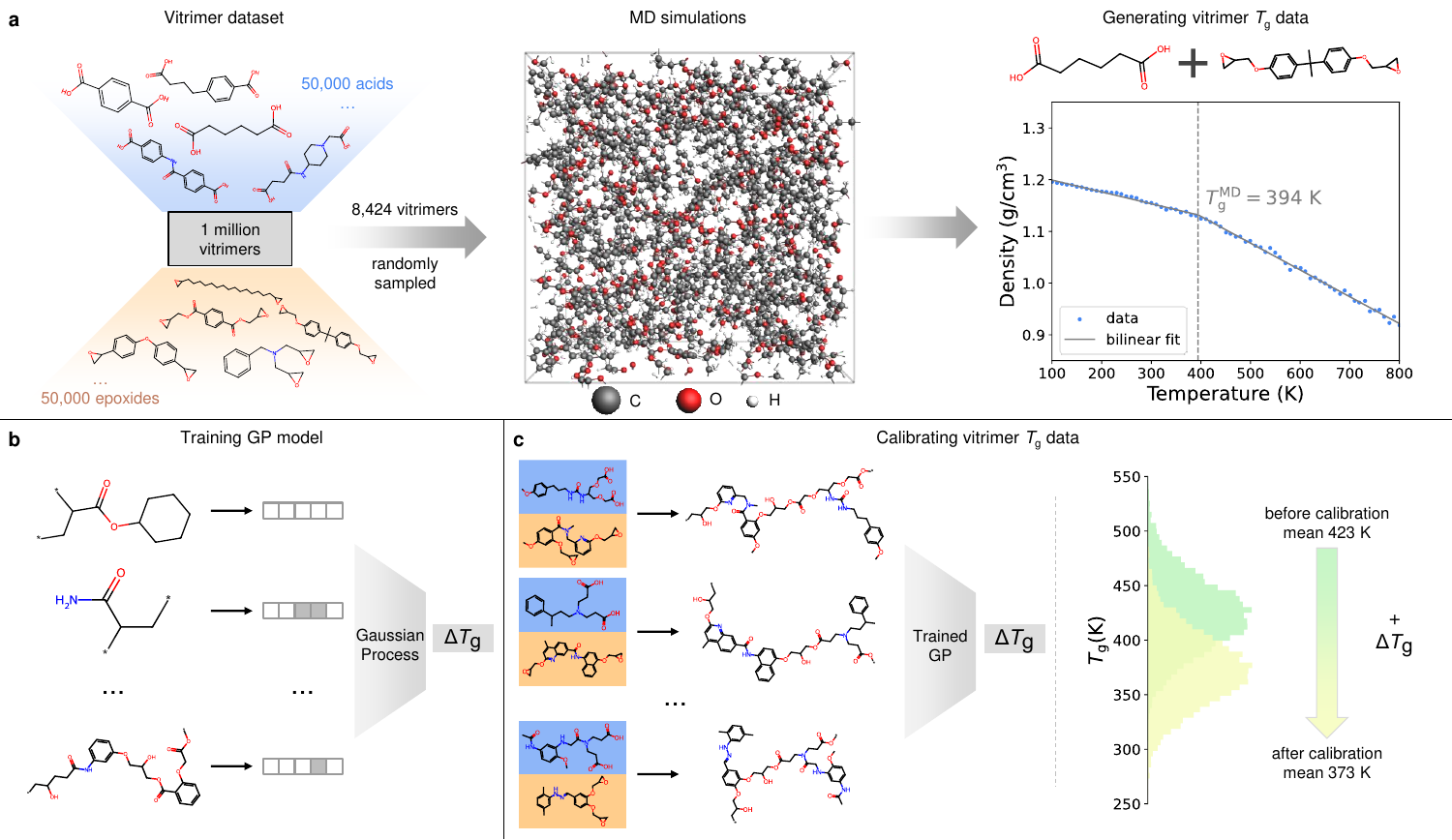}
\caption{\textbf{Data generation by MD simulations and calibration by GP model.} \textbf{(a)} The vitrimer dataset is obtained by randomly sampling one million combinations between 50,000 bifunctional carboxylic acids and 50,000 epoxides derived from the ZINC15 database. MD simulations are carried out to calculate $T_\mathrm{g}$ on a subset of 8,424 vitrimers. \textbf{(b)} We train a GP model to predict experiment-MD difference $\Delta T_\mathrm{g}$ with a training set of 295 polymers with experimental $T_\mathrm{g}$ in literature. \textbf{(c)} Using the trained GP model, we calibrate MD-calculated $T_\mathrm{g}$ of the vitrimer dataset. The calibrated $T_\mathrm{g}$, serving as a proxy of experimental $T_\mathrm{g}$, is the design target of this work.}
\label{data}
\end{figure}

We begin by creating a vitrimer dataset to train the VAE model. Since there are only a few available bifunctional transesterification vitrimers recorded in literature, we create a dataset of hypothetical vitrimers by combining carboxylic acids and epoxides. We first build two datasets by collecting available bifunctional carboxylic acids and epoxides from the online chemical compound database ZINC15 \cite{sterling2015zinc}. To further broaden the chemical space, we augment the datasets by adding hypothetical carboxylic acids and epoxides derived from available alcohols, olefins and phenols in the ZINC15 database. In both datasets, molecules satisfying all following rules are kept: (i) Carboxylic acid and epoxide-containing monomers have exactly two occurrences of their defining functional group (to restrict compositions to linear chains). (ii) Molecules with molecular weight smaller than 500 g/mol (to restrict the sizes of the molecular graphs and facilitate training of the graph VAE). (iii) Molecules with C, H, N, O elements only (to emulate the existing transesterification vitrimers). After filtering, two datasets of around 322,000 carboxylic acids and 625,000 epoxides are constructed. To ensure synthesizability, we select the 50,000 acids and 50,000 epoxides with lowest synthetic accessibility (SA) scores \cite{ertl2009estimation} (i.e., those predicted to be easiest to synthesize). The final dataset is built by randomly sampling one million vitrimers from the design space of 2.5 billion possible combinations between the selected acids and epoxides, as shown in Figure \ref{data}a.

To achieve property-guided inverse design, we further compute $T_\mathrm{g}$ of the vitrimers. Since MD simulations of the entire one-million dataset are computationally intractable, we calculate $T_\mathrm{g}$ of 8,424 vitrimers randomly sampled from the dataset. The quantity can cover a sufficient amount of vitrimers in the design space as well as keep the computational cost to a reasonable level. For each vitrimer, we create a virtual specimen then minimize and anneal the structure to remove local heterogeneities by slowly heating it to 800 K. A snapshot of the annealed system of an example vitrimer (adipic acid and bisphenol A diglycidyl ether) is shown in Figure \ref{data}a. The annealed structure is held at 800 K for an additional 50 ps and five specimens separated by 10 ps are obtained. To measure densities at different temperatures for $T_\mathrm{g}$ calculation, each specimen is cooled down from 800 K to 100 K linearly in steps of 10 K. By fitting a bilinear regression to the density-temperature profile, we calculate $T_\mathrm{g}$ as the intersection point of the two linear regressions (Figure \ref{data}a). Five replicate simulations are carried out from each specimen to reduce the noise due to the stochastic nature of MD. The distributions of average $T_\mathrm{g}$ and coefficient of variation (i.e., ratio of the standard deviation to the mean) in $T_\mathrm{g}$ of the vitrimers calculated by the five replicate MD simulations are shown in Supplementary Figures \ref{tghist}a and \ref{tghist}b, respectively. The coefficients of variation in $T_\mathrm{g}$ of most of the vitrimers are below 0.1 with only a few around 0.15, indicating the low uncertainty in our MD simulations. More details on MD simulations are provided in Supplementary Note \ref{supp_md}.

Due to the large difference in the cooling rate between MD simulations and experiments, MD-calculated $T_\mathrm{g}$ is typically overestimated compared with experimental measurements. Compensating for this artifact, Afzal et al. have achieved good correlation between MD-simulated $T_\mathrm{g}$ and experimental $T_\mathrm{g}$ on 315 polymers using ordinary least squares \cite{afzal2020high}. However, we find empirically that a simple two-parameter linear fit is insufficient to reduce the effect of larger noise in our MD simulations with smaller systems and fewer replicates (see Supplementary Figure \ref{calibration}c). Instead, we employ a Gaussian process (GP) regression model to calibrate MD calculations against available experimental data. GP is a probabilistic model that uses a kernel (covariance) function to make probabilistic predictions based on the distance between the queried data point and a training set \cite{deringer2021gaussian}. In order to construct a training dataset for the GP model, we gather 292 polymers from the Bicerano Handbook \cite{bicerano2002prediction} and the Chemical Retrieval on the Web (CROW) polymer database \cite{crow}, each with documented experimental $T_\mathrm{g}$. If the same polymer appears in both literature sources with different recorded $T_\mathrm{g}$, the average of both values is calculated as final $T_\mathrm{g}$. The available experimental data of three bifunctional transesterification vitrimers\cite{kamble2022reversing, ran2021dynamic, wu2021natural} is also included and the dataset contains 295 polymers in total. We compute $T_\mathrm{g}$ for this experimental polymer dataset using the MD protocols described above and calculate the experiment-MD difference $\Delta T_\mathrm{g}$ for each of these polymers. To numerically represent both the polymers within the training set and the vitrimers to be calibrated as inputs for the GP model, we apply extended-connectivity fingerprints (ECFPs) \cite{rogers2010extended} to the repeating units of the polymers, where asterisks (*) indicate connection points. We train the GP model to predict $\Delta T_\mathrm{g}$ from molecular fingerprints, as shown in Figure \ref{data}b. The calibrated $T_\mathrm{g}$ is defined as
\begin{equation}
    T_\mathrm{g} = T_\mathrm{g}^\mathrm{MD} + \Delta T_\mathrm{g}^\mathrm{GP},
\end{equation}
where $T_\mathrm{g}^\mathrm{MD}$ is the MD-calculated $T_\mathrm{g}$ and $\Delta T_\mathrm{g}^\mathrm{GP}$ is the experiment-MD $T_\mathrm{g}$ difference predicted by the GP model. More details on molecular fingerprints and the kernel function are provided in Supplementary Note \ref{supp_gp}.

To evaluate the performance of our GP model, we implement leave-one-out cross validation (LOOCV). In this process, we train our GP model on all data points in the training set except one point and predict its calibrated $T_\mathrm{g}$. We repeat this process for all 295 polymers in the training set and compare the calibrated $T_\mathrm{g}$ with experimental $T_\mathrm{g}$ recorded in literature. The mean absolute error (MAE) is 28.07 K (Supplementary Figure \ref{calibration}), which is comparable to the results by Afzal et al. \cite{afzal2020high} (MAE = 27.35 K) with larger systems and longer simulation time. The error is partially attributed to inconsistencies in recorded experimental $T_\mathrm{g}$ between the two literature sources \cite{bicerano2002prediction, crow}. Experimentally measured $T_\mathrm{g}$ values for some polymers (such as polypropylene, polyvinylcyclohexane, etc.) differ by approximately 30 K gathered from two literature sources \cite{bicerano2002prediction, crow}. By utilizing the comprehensive GP model trained on the entire training dataset without LOOCV, we proceed to calibrate $T_\mathrm{g}$ of the vitrimer dataset calculated by MD simulations. The distributions of $T_\mathrm{g}$ of the vitrimers before and after GP calibration are shown in Figure \ref{data}c and both distributions approach Gaussian. The average $T_\mathrm{g}$ before and after calibration is 423 K and 373 K, respectively. Since the cooling rate in our MD simulations is 12 orders of magnitude higher than typical experiments, the difference of 50 K is consistent with the Williams–Landel–Ferry theory that estimates an increase in $T_\mathrm{g}$ of 3 to 5 K per order of magnitude increase in the cooling rate \cite{ediger1996supercooled}. Ten vitrimers with highest and lowest calibrated $T_\mathrm{g}$ are shown in Supplementary Figure \ref{tg}, indicating the wide chemical and property space covered by the dataset. In this work, we denote $T_\mathrm{g}$ as the calibrated value from MD simulations, which serves as a proxy of the true experimental $T_\mathrm{g}$. It is also the input to the variational autoencoder and target of inverse design.

\subsection*{Variational autoencoder}

\begin{figure}[t]
\centering
\includegraphics[width=\linewidth]{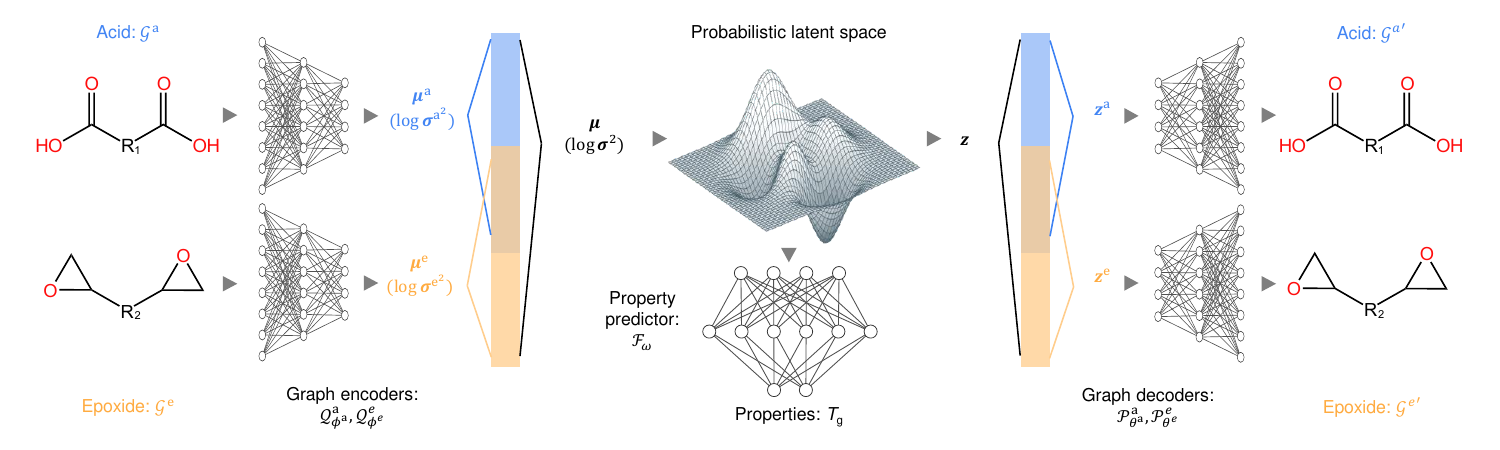}
\caption{\textbf{Illustration of the VAE model.} The encoders convert acid and epoxide molecules into latent vectors $\boldsymbol{z}$ in a continuous latent space. The latent vectors $\boldsymbol{z}$ are further decoded into acid and epoxide molecules by the decoders. A property predictor is added to predict $T_\mathrm{g}$ from $\boldsymbol{z}$.}
\label{vae}
\end{figure}

The discrete nature of molecules makes it challenging for the generative model to learn a continuous latent space from discrete data of vitrimers. Any two molecules can have different degrees of freedom (e.g., number of atoms and bonds) and extra attention needs to be paid to the choice of representations. Here we adopt the hierarchical graph representation of molecules developed by a Jin et al. \cite{jin2020hierarchical}. A molecule is first represented as a graph $\mathcal{G} = (\mathcal{V}, \mathcal{E})$ with atoms as nodes $\mathcal{V}$ and bonds as edges $\mathcal{E}$. We decompose the molecule $\mathcal{G}$ into $n$ motifs $\mathcal{M}_1, ..., \mathcal{M}_n$. Each motif $\mathcal{M}_i = (\mathcal{V}_i, \mathcal{E}_i)$ where $i \in \{1, ..., n\}$ is a subgraph with atoms $\mathcal{V}_i$ and edges $\mathcal{E}_i$. The ensuing step involves a three-level hierarchical graph representation (see Supplementary Figure \ref{hierarchical} for a schematic illustration). The motif level $\mathcal{G}_\mathcal{M}$ establishes macroscopic connections in a tree-like structure, the attachment level $\mathcal{G}_\mathcal{A}$ encodes inter-motif connectivity via shared atoms, and the atom level $\mathcal{G}$ captures finer atomic relationships. More details about the hierarchical graph representation are presented in Supplementary Note \ref{supp_repre}.

We use a variational autoencoder (VAE) comprising two pairs of hierarchical encoders and decoders associated with the hierarchical representations of acid and epoxide molecules, respectively. A schematic of the framework is presented in Figure \ref{vae}. Each of the hierarchical encoder uses three message passing networks (MPNs) to encode the graphs from each of the three levels. The acid encoder $\mathcal{Q}^\mathrm{a}_{\phi^\mathrm{a}}$ (with trainable parameters $\phi^\mathrm{a}$) maps the molecular graph of the acid molecule $\mathcal{G}^\mathrm{a}$ into a pair of vectors $\boldsymbol{\mu}^\mathrm{a} \in \mathbb{R}^{d_\mathrm{a}}$ and $\log {\boldsymbol{\sigma}^\mathrm{a}}^2 \in \mathbb{R}^{d_\mathrm{a}}$ of dimension $d_\mathrm{a}$, which are the mean and logarithm variance of a Gaussian distribution. Similarly, $\boldsymbol{\mu}^\mathrm{e} \in \mathbb{R}^{d_\mathrm{e}}$ and $\log {\boldsymbol{\sigma}^\mathrm{e}}^2 \in \mathbb{R}^{d_\mathrm{e}}$ of dimension $d_\mathrm{e}$ are converted from the epoxide molecule $\mathcal{G}^\mathrm{e}$ by the epoxide encoder $\mathcal{Q}^\mathrm{e}_{\phi^\mathrm{e}}$.

We employ the attributed network embedding method \cite{lerique2020joint, zheng2023unifying} to obtain the unified mean $\boldsymbol{\mu}$ and log variance $\log \boldsymbol{\sigma}^2$ of dimension $d$ embedding information of the acid and epoxide as well as their unified effects as follows. We define $d_\mathrm{ae} = d_\mathrm{a} + d_\mathrm{e} - d$ denoting the overlapping dimensions of $\boldsymbol{\mu}^\mathrm{a}$ and $\boldsymbol{\mu}^\mathrm{e}$ and calculate $\boldsymbol{\mu}$ by
\begin{equation} \label{partial}
    \boldsymbol{\mu} = \underbrace{\begin{bmatrix}
           \mu_{1}^\mathrm{a} \\
           \vdots \\
           \mu_{d_\mathrm{a}-d_\mathrm{ae}}^\mathrm{a}
         \end{bmatrix}}_{\text{acid-specific}} \oplus
         \underbrace{\frac{1}{2}
         \left(\begin{bmatrix}
         \mu_{d_\mathrm{a}-d_\mathrm{ae}+1}^\mathrm{a} \\
           \vdots \\
           \mu_{d_\mathrm{a}}^\mathrm{a}
         \end{bmatrix} + \begin{bmatrix}
             \mu_{1}^\mathrm{e} \\
           \vdots \\
           \mu_{d_\mathrm{ae}}^\mathrm{e}
         \end{bmatrix}
         \right)}_{\text{shared}} \oplus
          \underbrace{\begin{bmatrix}
           \mu_{d_{\mathrm{ae}+1}}^\mathrm{e} \\
           \vdots \\
           \mu_{d_\mathrm{e}}^\mathrm{e}
         \end{bmatrix}}_{\text{epoxide-specific}},
\end{equation}
where $\oplus$ denotes vector concatenation. The unified log variance vector $\log {\boldsymbol{\sigma}}^2$ is obtained similarly. Partially overlapping latent dimensions enable both independent and joint control as well as interpretability of embeddings of acid and epoxide. When exploring the latent space and optimizing latent vectors for new vitrimers later, it also allows us to change one part of the vitrimer but keep the other one unaltered.

The unified mean $\boldsymbol{\mu}$ and variance ${\boldsymbol{\sigma}}^2$ together describe a diagonal multivariate Gaussian distribution
\begin{equation}
    \boldsymbol{z} \sim \mathcal{N} \left(
    \begin{bmatrix}
        \mu_1, ..., \mu_d
    \end{bmatrix}^\intercal, \mathrm{diag} \left(
    \begin{bmatrix}
        \sigma_1^2, ..., \sigma_d^2
    \end{bmatrix}^\intercal
    \right)
    \right),
\end{equation}
where $\boldsymbol{z}$ is the latent vector (representation) of dimension $d$ encoding necessary information of both input graphs $\mathcal{G}^\mathrm{a}$ and $\mathcal{G}^\mathrm{e}$. To keep differentiability and facilitate the training of VAE, the reparameterization trick \cite{kingma2013auto} is used to sample the latent vector $\boldsymbol{z}$ from $\boldsymbol{\mu}$ and ${\boldsymbol{\sigma}}^2$ by
\begin{equation}
    \boldsymbol{z} = \boldsymbol{\mu} + \boldsymbol{\epsilon} \odot \begin{bmatrix}
        \sigma_1^2, ..., \sigma_d^2
    \end{bmatrix}^\intercal,
\end{equation}
where $\boldsymbol{\epsilon} \sim \mathcal{N}(\boldsymbol{0}, \boldsymbol{I})$ is a vector of dimension $d$ and $\odot$ denotes element-wise multiplication. The acid decoder $\mathcal{P}^\mathrm{a}_{\theta^\mathrm{a}}$ (with trainable parameters $\theta^\mathrm{a}$) is used to output the acid molecule ${\mathcal{G}^{\mathrm{a}}}^{\prime}$ from the acid-specific and shared dimensions of $\boldsymbol{z}$. Similarly, the epoxide decoder $\mathcal{P}^\mathrm{e}_{\theta^\mathrm{e}}$ is used to output the epoxide molecule ${\mathcal{G}^{\mathrm{e}}}^{\prime}$ from the epoxide-specific and shared dimensions of $\boldsymbol{z}$. More specifically, the decoders iteratively expand the graphs at three hierarchical levels. At step $t$, three multilayer perceptrons (MLPs) are used to predict the probability distributions of each motif node $\boldsymbol{p}_{\mathcal{M}_t}^{\prime}$, attachment node $\boldsymbol{p}_{\mathcal{A}_t}^{\prime}$ and atoms to be attached $\boldsymbol{p}_{(u, v)_t}^{\prime}$ (see Supplementary Note \ref{supp_vae} for more details). An additional MLP is used to predict the probability of backtracing $\boldsymbol{p}_{\mathrm{b}_t}^{\prime}$, i.e., when there will be no new neighbors to add to the motif node. Both decoders are optimized to accurately reconstruct the molecules, i.e., ${\mathcal{G}^{\mathrm{a}}}^{\prime} \approx \mathcal{G}^\mathrm{a}$ and ${\mathcal{G}^{\mathrm{e}}}^{\prime} \approx \mathcal{G}^\mathrm{e}$. Practically this is achieved by minimizing the error between all four predicted probability distributions with respect to the one-hot encoded ground truth, i.e., $\boldsymbol{p}_{\mathcal{M}_t}$, $\boldsymbol{p}_{\mathcal{A}_t}$, $\boldsymbol{p}_{(u, v)_t}$ and $\boldsymbol{p}_{\mathrm{b}_t}$ for $t = 1, ..., t_\mathrm{max}$ where $t_\mathrm{max}$ is the maximum number of iterations based on depth-first search of the input molecule (here for simplicity we omit superscripts a and e denoting acid and epoxide). Encoding input vitrimers as described here introduces an information bottleneck \cite{tishby2000information} within the latent representation. This bottleneck selectively retains necessary information required for accurate vitrimer reconstruction while largely reducing the dimensionality and complexity of original data. As a result, vitrimers with similar compositions occupy proximate positions in the latent space.

In order to achieve data-driven design and uncover novel vitrimers with the interested property, we establish a connection between the latent space and $T_\mathrm{g}$. This is accomplished by employing a neural network surrogate model that takes the latent vectors as inputs and outputs $T_\mathrm{g}$. Consequently, we modify the original VAE architecture and establish a projection from the latent space to $T_\mathrm{g}$ by incorporating the latent vectors $\boldsymbol{z}$ into a property prediction model $\mathcal{F}_\omega$ (with trainable parameters $\omega)$. Thereby, the predicted property is 
\begin{equation}
    T_\mathrm{g}^{\prime} = \mathcal{F}_\omega(\boldsymbol{z}).
\end{equation}

We collect two subsets from the vitrimer dataset, one with $N$ vitrimers lacking property labels $\mathcal{D} = \{({\mathcal{G}^\mathrm{a}}^{(i)}, {\mathcal{G}^\mathrm{e}}^{(i)}): i = 1, ..., N\}$, and one with $N_\mathrm{prop}$ vitrimer and $T_\mathrm{g}$ pairs $\mathcal{D}_\mathrm{prop} = \{({\mathcal{G}^\mathrm{a}}^{(i)}, {\mathcal{G}^\mathrm{e}}^{(i)}, T_\mathrm{g}^{(i)}): i = 1, ..., N_\mathrm{prop}\}$. Due to the large difference between $N$ and $N_\mathrm{prop}$ (999,000 vs. 7,424), we first train the VAE on an unsupervised basis with $\mathcal{D}$ and the property predictor is not optimized. Specifically, 
\begin{align} \label{loss1}
    \theta^\mathrm{a}, \theta^\mathrm{e}, \phi^\mathrm{a}, \phi^\mathrm{e} \leftarrow \argmin_{\theta^\mathrm{a}, \theta^\mathrm{e}, \phi^\mathrm{a}, \phi^\mathrm{e}} 
    &\underbrace{\sum_{i=1}^{N} \left( \mathrm{CE}\left( {\boldsymbol{p}_{\mathcal{M}}^{(i)}}^{\prime}, \boldsymbol{p}_{\mathcal{M}}^{(i)}\right) + \mathrm{CE}\left( {\boldsymbol{p}_{\mathcal{A}}^{(i)}}^{\prime}, \boldsymbol{p}_{\mathcal{A}}^{(i)}\right) + \mathrm{CE}\left( {\boldsymbol{p}_{(u, v)}^{(i)}}^{\prime}, \boldsymbol{p}_{(u, v)}^{(i)}\right) + \mathrm{BCE}\left( {\boldsymbol{p}_\mathrm{b}^{(i)}}^{\prime}, \boldsymbol{p}_\mathrm{b}^{(i)}\right) \right)}_\text{reconstruction loss} \nonumber \\
    + &\underbrace{\lambda_\mathrm{KL} \frac{1}{N} \sum_{i=1}^{N} D_\mathrm{KL} \left( \mathcal{N} \left( \begin{bmatrix}
        \mu_1^{(i)}, ..., \mu_d^{(i)}
    \end{bmatrix}^\intercal, \mathrm{diag}\left( \begin{bmatrix}
        {\sigma_1^{(i)}}^2, ..., {\sigma_d^{(i)}}^2
    \end{bmatrix}^\intercal \right) \right) \|~\mathcal{N} \left(\boldsymbol{0}, \boldsymbol{I} \right) \right)}_\text{Kullback–Leibler divergence},
\end{align}
where CE and BCE denote cross entropy loss and binary cross entropy loss \cite{good1952rational}, respectively. For simplicity, the subscript $t$ and superscripts a and e are omitted and all terms in reconstruction loss represent the sum over all decoding steps and over acid and epoxide. $\lambda_\mathrm{KL} > 0$ is the regularization weight for Kullback–Leibler divergence. Training the VAE with $\mathcal{D}$ aims to construct well-trained encoders and decoders capable of accommodating a diverse array of vitrimers. The reconstruction loss ensures the accurate reconstruction of the encoded vitrimers with respect to both acid and epoxide molecules by the VAE. The Kullback–Leibler divergence (KLD)\cite{kullback1951information} is a statistical measure to quantify how different two distributions are from each other. Hence, by employing it as a loss term\cite{kingma2013auto}, we minimize the difference between the probability distribution of the latent space created by the encoder and the standard Gaussian distribution $\mathcal{N}(\boldsymbol{0}, \boldsymbol{I})$. This helps in constructing a seamless and continuous latent space from which new samples can be generated using standard Gaussian distribution and allows us to discover and design novel vitrimers not present in the training set. The KLD is calculated as
\begin{equation}
    D_\mathrm{KL} \left( \mathcal{N} \left( \begin{bmatrix}
        \mu_1, ..., \mu_d
    \end{bmatrix}^\intercal, \mathrm{diag}\left( \begin{bmatrix}
        \sigma_1^2, ..., \sigma_d^2
    \end{bmatrix}^\intercal \right) \right) \|~\mathcal{N} \left(\boldsymbol{0}, \boldsymbol{I} \right) \right) = \frac{1}{2} \sum_{j=1}^d [\mu_j^2 + \sigma_j^2 - \log (\sigma_j^2) - 1].
\end{equation}

Subsequently, we use $\mathcal{D}_\mathrm{prop}$ to jointly train encoder, decoder and property predictor at the same time, i.e., 
\begin{align} \label{loss2}
    \theta^\mathrm{a}, \theta^\mathrm{e}, \phi^\mathrm{a}, \phi^\mathrm{e}, {\color{red} \omega} \leftarrow \argmin_{\theta^\mathrm{a}, \theta^\mathrm{e}, \phi^\mathrm{a}, \phi^\mathrm{e}, {\color{red} \omega}} 
    &\underbrace{\sum_{i=1}^{N_\mathrm{prop}} \left( \mathrm{CE}\left( {\boldsymbol{p}_{\mathcal{M}}^{(i)}}^{\prime}, \boldsymbol{p}_{\mathcal{M}}^{(i)}\right) + \mathrm{CE}\left( {\boldsymbol{p}_{\mathcal{A}}^{(i)}}^{\prime}, \boldsymbol{p}_{\mathcal{A}}^{(i)}\right) + \mathrm{CE}\left( {\boldsymbol{p}_{(u, v)}^{(i)}}^{\prime}, \boldsymbol{p}_{(u, v)}^{(i)}\right) + \mathrm{BCE}\left( {\boldsymbol{p}_\mathrm{b}^{(i)}}^{\prime}, \boldsymbol{p}_\mathrm{b}^{(i)}\right) \right)}_\text{reconstruction loss} \nonumber \\
    + &\underbrace{\lambda_\mathrm{KL} \frac{1}{N_\mathrm{prop}} \sum_{i=1}^{N_\mathrm{prop}} D_\mathrm{KL} \left( \mathcal{N} \left( \begin{bmatrix}
        \mu_1^{(i)}, ..., \mu_d^{(i)}
    \end{bmatrix}^\intercal, \mathrm{diag}\left( \begin{bmatrix}
        {\sigma_1^{(i)}}^2, ..., {\sigma_d^{(i)}}^2
    \end{bmatrix}^\intercal \right) \right) \|~\mathcal{N} \left(\boldsymbol{0}, \boldsymbol{I} \right) \right)}_\text{Kullback–Leibler divergence} \nonumber \\
    + &
    \begingroup
    \color{red}
    \underbrace{\color{black} \frac{1}{N_\mathrm{prop}} \sum_{i=1}^{N_\mathrm{prop}} \left( {T_\mathrm{g}^{(i)}}^{\prime} - T_\mathrm{g}^{(i)}\right)^2}_\text{property prediction loss}
    \endgroup.
\end{align}
The additional property prediction loss ensures accurate prediction of $T_\mathrm{g}$ from latent vectors. This joint training process reorganizes the latent space and places vitrimers with similar $T_\mathrm{g}$ in close proximity to each other. More details about hierarchical encoder and decoder, network architecture, training protocols and hyperparameters are presented in Supplementary Notes \ref{supp_repre} and \ref{supp_vae}.

\subsection*{Performance of the VAE}

We first evaluate the ability of the VAE to reconstruct a given vitrimer. We encode the vitrimers in the test set into mean vectors of latent distribution $\boldsymbol{\mu}$ then decode $\boldsymbol{\mu}$ back to vitrimers. The ratio of successfully reconstructed (i.e., both carboxylic acid and epoxide decoded from $\boldsymbol{\mu}$ are identical to input molecules) is 85.4\%, which demonstrates well-trained encoders and decoders capable of accommodating and reconstructing vitrimers unseen by the VAE. Examples of ten vitrimers from the test set and the corresponding reconstructions are presented in Supplementary Figure \ref{reconstruct}. Eight vitrimers are perfectly reconstructed, while one component of vitrimers is decoded into different but similar molecules in the two unsuccessful examples.

We then assess the performance of the VAE to generate vitrimers. We sample 1,000 latent vectors $\boldsymbol{z}$ from standard Gaussian distribution and decode them into the carboxylic acid and epoxide molecules constituting vitrimers. 85.2\% of the sampled vitrimers are valid, i.e., the composing acid and epoxide molecules are chemically valid and contain exactly two carboxylic acid and epoxide groups. Even though the decoders are not explicitly coded to only output carboxylic acids and epoxides (e.g., by making the decoders build up molecules based on two carboxylic acid and epoxide functional groups), most of the randomly sampled latent vectors are decoded to molecules containing the desired functionality. Examples of sampled vitrimers are shown in Supplementary Figure \ref{sample}. Components of the three invalid sampled vitrimers are also carboxylic acids and epoxides but do not have exactly two functional groups.

Apart from validity, we are also interested in novelty and uniqueness of the generated vitrimers, which are defined as the ratio of sampled vitrimers which are not present in the training set and the expected fraction of unique vitrimers per sampled vitrimers, respectively. Results show that all of the 1,000 vitrimers sampled from the latent space are novel and unique, which greatly benefits the discovery of vitrimers by exploring the latent space.

We further examine the effect of joint training with the small dataset $\mathcal{D}_\mathrm{prop}$ containing a limited number of labeled vitrimers. All four metrics of the model before and after joint training are presented in Supplementary Table \ref{metrics}. The improved reconstruction accuracy and sample validity show that the second-step joint training enhances the performance of the model and that the encoders and decoders are not biased to the limited data in $\mathcal{D}_\mathrm{prop}$.

The property predictor maps latent space encoded from vitrimers to $T_\mathrm{g}$ and serves as a surrogate model for estimating $T_\mathrm{g}$ without the need for costly MD simulations. We evaluate the predictive power of the property predictor network by encoding the vitrimers in the test set into mean vectors $\boldsymbol{\mu}$ and predicting the associated $T_\mathrm{g}$. The predicted $T_\mathrm{g}$ and true $T_\mathrm{g}$ are compared in Supplementary Figure \ref{prediction}. A mean absolute error of 13.53 K indicates accurate prediction of $T_\mathrm{g}$ by the property predictor which facilitates the inverse design process.

The VAE jointly trained with the property predictor organizes the latent space such that vitrimers exhibiting similar properties are positioned in the vicinity of each other. We examine the distribution of latent vectors and corresponding $T_\mathrm{g}$ of the labeled datasets using principal component analysis (PCA). As shown in Supplementary Figure \ref{latent}a and \ref{latent}b, the distribution of the latent vectors shows an obvious gradient in both training and test sets, where vitrimers with higher $T_\mathrm{g}$ cluster in the lower left region. Such a well-structured latent space based on properties benefits discovery of novel vitrimers with desired $T_\mathrm{g}$. For comparison, the latent vector distribution before joint training is presented in Supplementary Figure \ref{latent}c and \ref{latent}d. The much less obvious trend confirms the effect of joint training on latent space organization.

\subsection*{Interpretable exploration of the latent space}

\begin{figure}
\centering
\includegraphics[width=\linewidth]{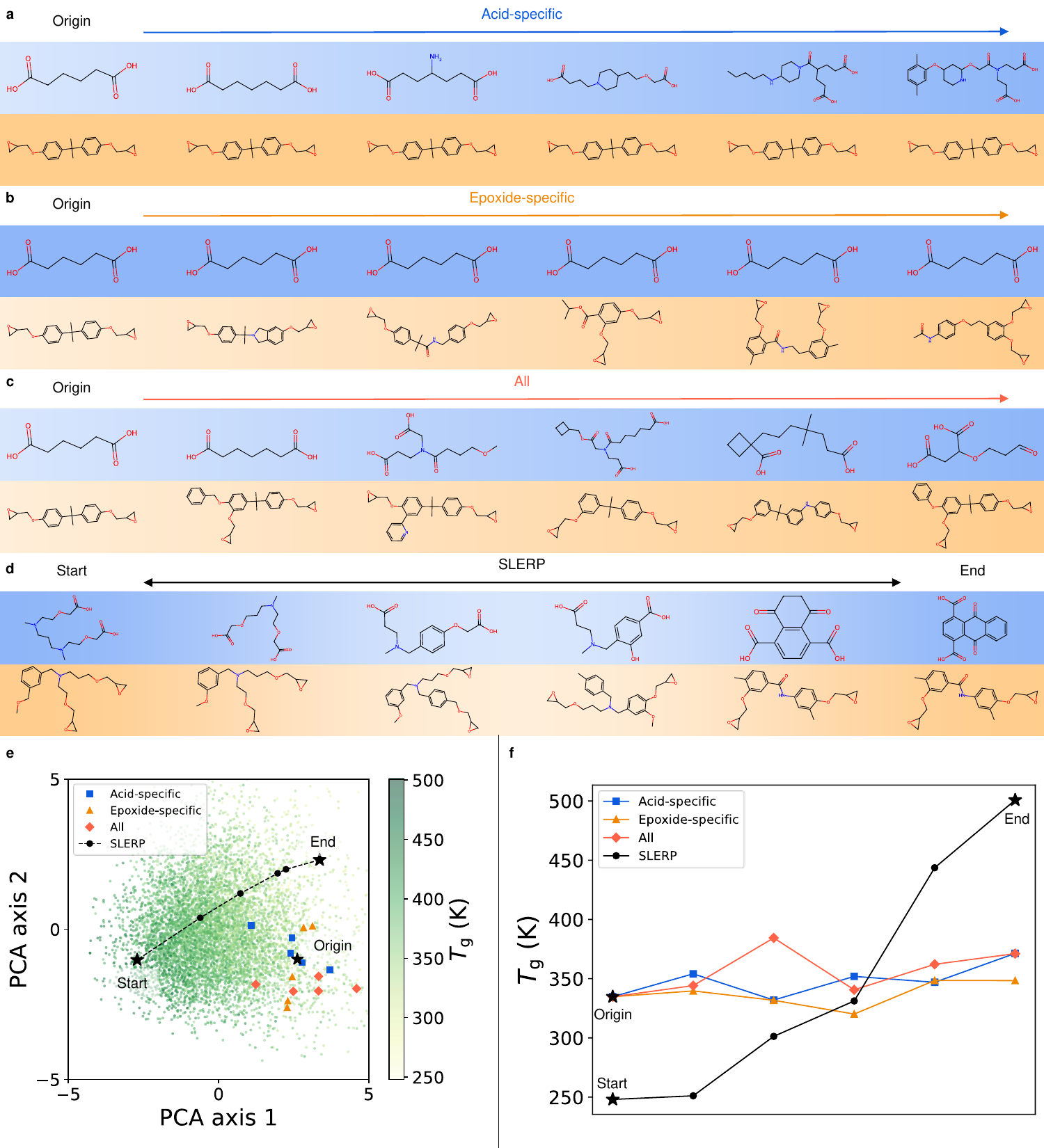}
\caption{\textbf{Exploration in the latent space to discover novel vitrimers.} \textbf{(a)(b)(c)} Starting with a known vitrimer as origin (adipic acid and bisphenol A diglycidyl ether), vitrimers are discovered by perturbing its latent vector in \textbf{(a)} acid-specific dimensions, \textbf{(b)} epoxide-specific dimensions and \textbf{(c)} all dimensions. \textbf{(d)} Novel vitrimers are identified along the interpolation path between two vitrimers in the training set. \textbf{(e)} The distribution of discovered vitrimers is visualized in the latent space by PCA. \textbf{(f)} $T_\mathrm{g}$ of discovered vitrimers. All presented $T_\mathrm{g}$ values of are validated by MD simulations and GP calibration.}
\label{explore}
\end{figure}

The well trained, continuous latent space enables us to discover new vitrimers by exploring the latent space through modifications of latent vectors $\boldsymbol{z}$. For example, we start with the latent vector $\boldsymbol{z}_0$ of a known vitrimer (adipic acid and bisphenol A diglycidyl ether) as origin and sample latent vectors in the neighborhood by perturbing $\boldsymbol{z}_0$. Previous works that employ multi-component VAEs (i.e., VAEs with multiple encoders and decoders) simply add embedding or mean (log variance) vectors from encoders to derive the unified latent vector $\boldsymbol{z}$ \cite{yao2021inverse}. The effect of different components is not considered individually and a change in $\boldsymbol{z}$ leads to potential changes in all components. The partially overlapping latent dimensions (Equation \ref{partial}) allow us to explore the vicinity of the origin $\boldsymbol{z}_0$ along different axes by adding noise to acid-specific latent dimensions (first $d_\mathrm{a}$ dimensions of $\boldsymbol{z}_0$), epoxide-specific latent dimensions (last $d_\mathrm{e}$ dimensions of $\boldsymbol{z}_0$) and all latent dimensions of $\boldsymbol{z}_0$ (details are provided in Supplementary Note \ref{supp_explore}). Consequently, novel vitrimers with changes in only acid, only epoxide and both components are identified by decoding the latent vectors modified along three axes, as shown in Figures \ref{explore}a-c. The decoded vitrimers present variety in molecular structures without significant changes in $T_\mathrm{g}$ (Figure \ref{explore}f) due to limited search region in latent space, which opens an opportunity to tailor a specific vitrimer to its novel variants with different molecular structures but preserve certain property similarity.

Besides neighborhood search, we perform an interpolation between two points in the latent space and identify a series of new vitrimers along the path. Figure \ref{explore}d presents an example of spherical interpolation (SLERP) \cite{shoemake1985animating} between vitrimers with highest and lowest $T_\mathrm{g}$ in the training set. As opposed to linear interpolation (LERP), we use SLERP because Gaussian distribution in high dimensions closely follows the surface of a hypersphere. The decoded vitrimers show a smooth transition from the low-$T_\mathrm{g}$ vitrimer with linear structure to the high-$T_\mathrm{g}$ vitrimer with more aromatic nature. The continuous transition between molecular structures and $T_\mathrm{g}$ (Figure \ref{explore}f) evidences the smoothness of the latent space with associated $T_\mathrm{g}$. The vitrimers discovered by LERP and their associated $T_\mathrm{g}$ are shown in Supplementary Figure \ref{inter}. More details on spherical and linear interpolation schemes are presented in Supplementary Note \ref{supp_explore}.

\subsection*{Inverse design by Bayesian optimization}

\begin{figure}
\centering
\includegraphics[width=\linewidth]{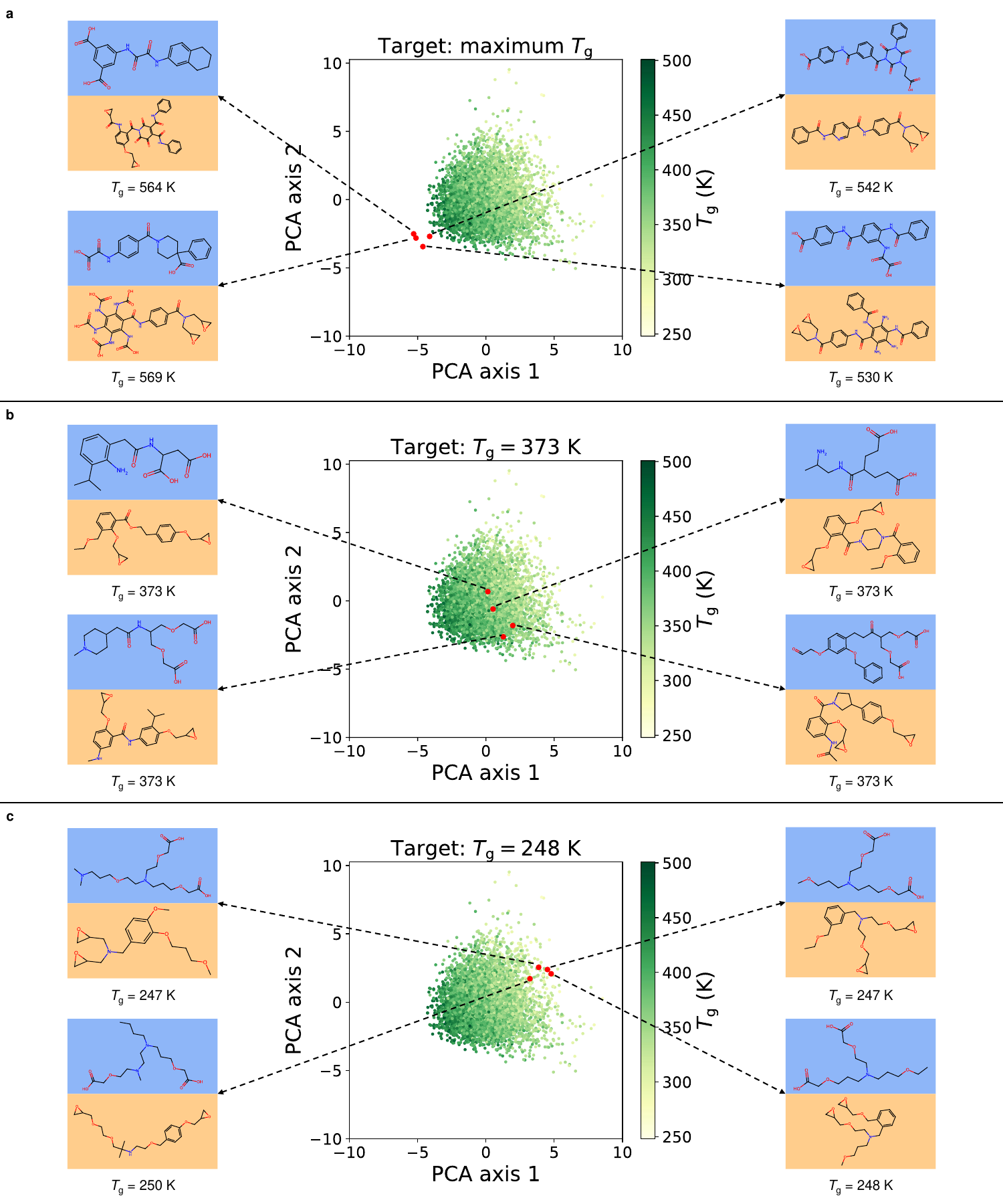}
\caption{\textbf{Inverse design of novel vitrimers by Bayesian optimization based on three targets of desirable $T_\mathrm{g}$:} \textbf{(a)} Maximum $T_\mathrm{g}$, \textbf{(b)} $T_\mathrm{g} = 373~\mathrm{K}$ and \textbf{(c)} $T_\mathrm{g} = 248~\mathrm{K}$. All presented $T_\mathrm{g}$ values of proposed vitrimers are validated by MD simulations and GP calibration.}
\label{bo}
\end{figure}

The VAE together with the property predictor succeeds in learning the hidden relationships between latent space and $T_\mathrm{g}$ of vitrimers, which allows us to tailor vitrimer compositions to desirable $T_\mathrm{g}$ even beyond the training regime. Although we have achieved forward projection from the vitrimer space (or latent space) to property space, the inverse mapping is more challenging due to the fact that multiple distinct vitrimers could have a similar $T_\mathrm{g}$. To achieve inverse design of vitrimers with optimal or desirable $T_\mathrm{g}$, we employ batch Bayesian optimization to identify the latent vectors $\boldsymbol{z}$ that have the potential to be decoded into vitrimers with target $T_\mathrm{g}$. The proposed candidates are further validated by MD simulations with GP calibration, and the optimal solutions with desirable $T_\mathrm{g}$ are found. Due to the discrete nature of molecules, the latent vectors proposed by the optimization process may lead to invalid molecules. Furthermore, since the discrete molecules are projected onto a continuous latent space by the VAE, it is inevitable that multiple distinct latent vectors in the neighborhood can be decoded into the same vitrimer but are associated with different $T_\mathrm{g}$ predicted by the property predictor. This severely limits the accuracy and efficiency of the optimization process. To this end, we add an additional decoding-encoding step before passing $\boldsymbol{z}$ to the property predictor to predict $T_\mathrm{g}$ (Supplementary Figure \ref{boflow}). More specifically, when evaluating the $T_\mathrm{g}$ of a point of interest $\boldsymbol{z}$ in the latent space during the optimization process, $\boldsymbol{z}$ is first decoded into a carboxylic acid and an epoxide. If both molecules are valid, they are passed to the encoders to obtain the reconstructed mean vector $\boldsymbol{\mu}_\mathrm{recon}$, which is further passed to the property predictor to evaluate the $T_\mathrm{g}$. In this way, the Bayesian optimization algorithm is able to search for potential candidates with desirable $T_\mathrm{g}$ efficiently without proposing the same vitrimer for a large number of iterations. More details about Bayesian optimization are provided in Supplementary Note \ref{supp_bo}.

To demonstrate the effectiveness of our inverse design framework, we use Bayesian optimization to discover novel vitrimers with three different targets: maximum $T_\mathrm{g}$, $T_\mathrm{g} = 373~\mathrm{K}$ and $T_\mathrm{g} = 248~\mathrm{K}$. $T_\mathrm{g}$ of the proposed candidates is validated by MD simulations and GP calibration. For each target, four examples of discovered vitrimers are presented in Figure \ref{bo}. For the first target (maximum $T_\mathrm{g}$), our VAE model generates novel vitrimers with MD-validated $T_\mathrm{g}$ beyond the upper bound of $T_\mathrm{g}$ in the training data (500 K) and thereby expands the limits in thermal properties of bifunctional transesterification vitrimers. The Bayesian optimization procedures are able to probe the latent space outside of the training domain and propose novel vitrimers with extreme properties, which is difficult for traditional forward modeling methods to find. For the second target, the Bayesian optimization algorithm effectively searches the latent space and successfully proposes vitrimers with the exact target $T_\mathrm{g}$ of 373 K. The corresponding latent vectors are spread out in the latent space and the vitrimer compositions present significant molecular variety. For the third target which is the lower bound of the training domain (248 K), the proposed vitrimers (especially carboxylic acids) are more similar to each other and occupy a small region in the latent space. This can be attributed to the fact that there are not many linear molecules with more aliphatic nature in the 50,000 acids or epoxides making the training set. As a result, the distribution of these vitrimers with low $T_\mathrm{g}$ is insufficiently captured by the VAE and the proposed candidates of low-$T_\mathrm{g}$ vitrimers are restrained by the limited training data.

The $T_\mathrm{g}$ distributions of the dataset and around 100 designed vitrimers for each target are presented in Supplementary Figure \ref{violin}. For the target of maximum $T_\mathrm{g}$, our model efficiently discovers novel vitrimer chemistries beyond the training property space. For the other two targets of 373 K and 248 K, the distributions are centered around the design target. Ten examples of novel vitrimers discovered by Bayesian optimization for each target are presented in Supplementary Figure \ref{morebo}. For the first target, all ten proposed vitrimers have validated $T_\mathrm{g}$ larger than 500 K, which indicates the effective extrapolation beyond the training domain by our framework. For the other two targets of finding vitrimers with exact target $T_\mathrm{g}$, the discovered vitrimers present $T_\mathrm{g}$ within a range of 2 K around the target and maintain considerable molecular diversity, proving the high accuracy in the inverse design process. We further examine the stability of the proposed molecules by minimizing them by reactive molecular dynamics (ReaxFF) using the CHON2017\_weak\_bb force field \cite{vashisth2018accelerated}. All molecules remain stable during minimization and the minimized structures are presented in Supplementary Figure \ref{minimized}.

We carry out further analysis based on the molecular descriptors of ten proposed vitrimers for each target. The molecular descriptors except density are calculated from the vitrimer repeating units (Supplementary Figure \ref{md}a with $n = 1$) by the Modred package \cite{moriwaki2018mordred}. Density of each vitrimer at 300 K is extracted from MD simulations. The relevant descriptors of designed low, medium and high temperature vitrimers are presented in Supplementary Figure \ref{descriptor}. The vitrimers with higher $T_\mathrm{g}$ have larger molecular weight, higher density, more heavy atoms and multiple bonds and fewer rotatable bonds. Consequently, the chains in these vitrimers are more rigid and less mobile, which agrees with the common knowledge of structure-$T_\mathrm{g}$ relationships in polymers.

We compare $T_\mathrm{g}$ of the designed vitrimers with nine commonly used polymers in Supplementary Figure \ref{tgbarchart}. The proposed vitrimers cover a wide range of $T_\mathrm{g}$ suitable for various applications from coating materials to aerospace polymers. With further tuning of the target, our framework has the potential to discover vitrimer compositions with any $T_\mathrm{g}$ within an expanded range and expedite the widespread applications of sustainable polymers in various industries.

\subsection*{Experimental synthesis of novel vitrimer designed with chemical intuition}

\begin{figure}
\centering
\includegraphics[width=\linewidth]{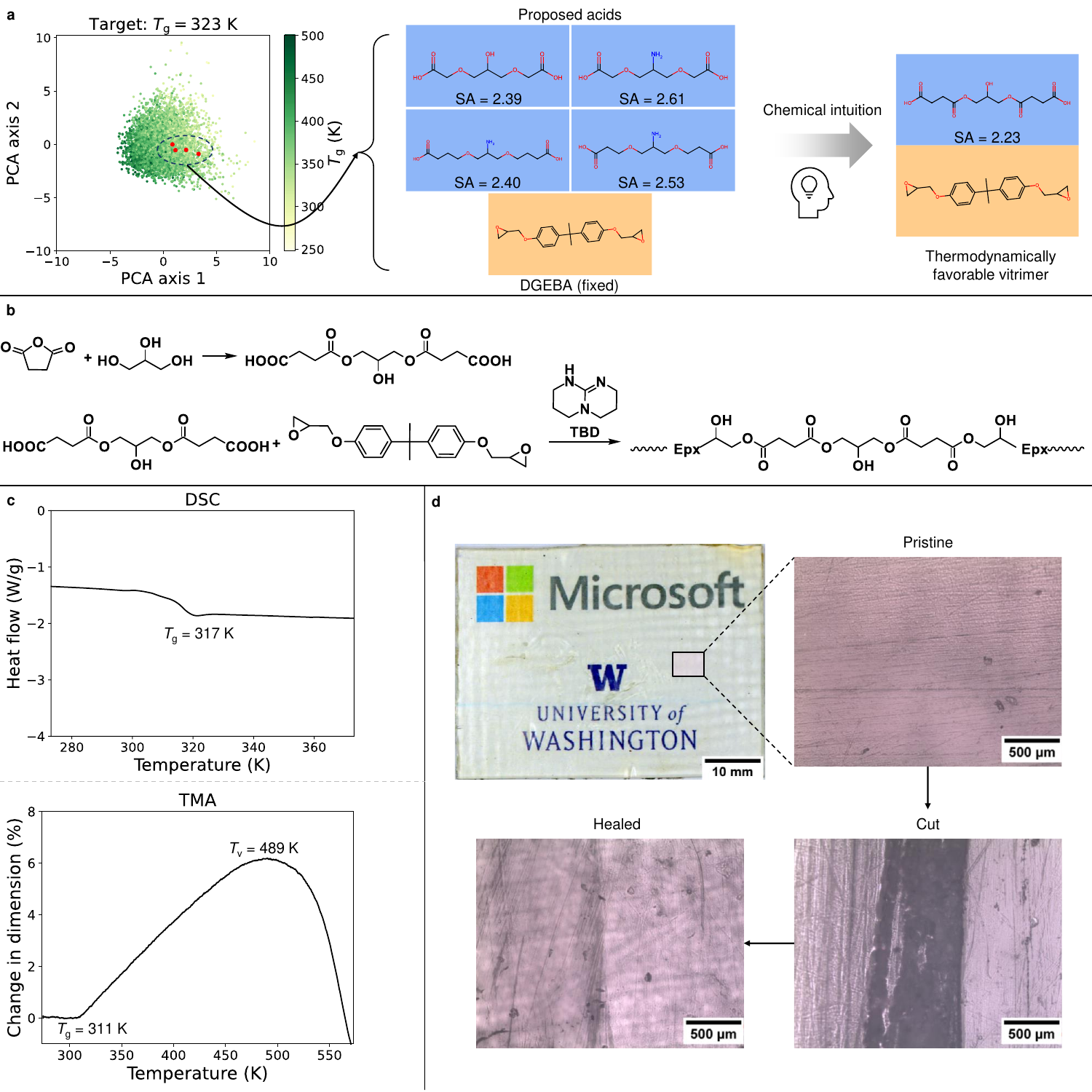}
\caption{\textbf{Synthesis and characterization of novel vitrimer designed by ML framework.} \textbf{(a)} Four vitrimers (epoxide fixed as DGEBA) are proposed by Bayesian optimization with a target $T_\mathrm{g} = 323~\mathrm{K}$. Driven by chemical intuition, a symmetric and thermodynamically stable vitrimer are selected. \textbf{(b)} Reaction scheme for synthesis of the acid and consequent crosslinking with DGEBA to form a novel vitrimer. \textbf{(c)} Experimental characterization of the synthesized vitrimer to measure $T_\mathrm{g}$ and $T_\mathrm{v}$. The measured $T_\mathrm{g}$ ranges from 311 K to 317 K, which agrees well with the design target. \textbf{(d)} Images of pristine, cut and healed vitrimer specimens, confirming healability of the synthesized vitrimer.}
\label{synthesis}
\end{figure}

To experimentally validate the effectiveness of the VAE model, we perform Bayesian optimization to propose novel vitrimers with a target $T_\mathrm{g}$ of 323 K. Since epoxides are typically more difficult to synthesize, the epoxide molecule is fixed as bisphenol A diglycidyl ether (DGEBA) during optimization to improve synthesizability of the vitrimer. In other words, the shared and epoxide-specific dimensions are fixed while we only optimize acid-specific dimensions in Equation \ref{partial}. Out of the carboxylic acids proposed by the VAE model, four acids with low SA scores (2.39 to 2.61) and symmetric structures are further analyzed for synthesis feasibility. Additionally, some of these acid molecules have an amine group that can react with epoxide rings to form irreversible covalent bonds, thereby reducing the adaptive nature of the macromolecular network. Keeping these thermodynamical stability issues in polymer synthesis and resulting crosslinked network through chemical intuition, we make slight modifications to the proposed structures to achieve a symmetric acid molecule with a lower SA score (2.23) that can be synthesized using off-the-shelf chemicals. This acid is crosslinked with DGEBA epoxide to form a stable polymer, demonstrating inverse design and synthesis of novel vitrimer chemsitry (Figure \ref{synthesis}a).

The synthesis of the novel vitrimer chemistry is carried out by the ring opening reaction of succinic anhydride by glycerol followed by immediate crosslinking with DGEBA. The reaction of beta-hydroxy groups of glycerol with succinic anhydride opens the ring and creates carboxylic functional groups. The resultant acid is crosslinked with DGEBA in presence of catalyst triazabicyclodecene (TBD) to yield the final vitrimer product (Figure \ref{synthesis}b). The cured vitrimer is verified using Fourier transform infrared (FTIR) spectroscopy, as shown in Supplementary Figure \ref{FTIR}. The characteristic peaks for carbonyl groups at 1728 $\mathrm{cm}^{-1}$ and for aromatic epoxide chains at 1033 to 1028 $\mathrm{cm}^{-1}$ are observed. In addition, other peaks at 1400 to 1608 $\mathrm{cm}^{-1}$ indicate the formation of ester linkages between acids and epoxides. We further characterize the thermal properties of the synthesized vitrimer using differential scanning calorimetry (DSC) and thermomechanical analysis (TMA), as shown in Figure \ref{synthesis}c. DSC result shows one transition temperature ($T_\mathrm{g}$) at $317~\mathrm{K}$, indicating complete curing of crosslinked vitrimer. TMA result presents two thermal transitions, first at $311~\mathrm{K}$ for $T_\mathrm{g}$ and second at $489~\mathrm{K}$ for $T_\mathrm{v}$. The difference in $T_\mathrm{g}$ from DSC and TMA arises due to different physical phenomena and sensitivities measured by each technique. DSC measures heat flow associated with the glass transition, which reflects changes in heat capacity as the polymer transitions from a glassy to a rubbery state, it therefore provides an average $T_\mathrm{g}$. TMA on the other hand measures the mechanical response of the polymer to non-isothermal creep and the machine detects subtle changes in molecular mobility through dimensional changes. Experimental $T_\mathrm{g}$ from DSC and TMA agrees well with the design target ($323~\mathrm{K}$) and demonstrates the capability of our framework to design novel vitrimers with desired experimental $T_\mathrm{g}$. The second transition indicates flowability of the vitrimer at elevated temperature when dynamic exchange reactions start to occur and enhance polymer chain mobility. To confirm healability of the vitrimer, we cut a pristine specimen and heal it at temperature around $T_\mathrm{v}$. The surfaces of pristine, cut, and healed samples are examined under the microscope (Figure \ref{synthesis}d). The complete removal of damage shows  healability and recyclability of the synthesized novel vitrimer chemistry.

\section*{Conclusion}

We develop an integrated MD-ML framework for inverse design of bifunctional transesterification vitrimers with desirable $T_\mathrm{g}$. A diverse vitrimer dataset is built for the first time from the ZINC15 database \cite{sterling2015zinc}. High-throughput MD simulations with a GP calibration model are employed to calculate $T_\mathrm{g}$ on a subset of vitrimers. The dataset is used to train a VAE model with dual graph encoders and decoders which enables representation and design of the desired vitrimer components. This further provides flexibility by exploring the latent space and optimizing latent vectors of different components for novel vitrimers. We demonstrate the high accuracy and efficiency of our framework in discovering novel vitrimers with three different targets of $T_\mathrm{g}$ even beyond the training distribution. The proposed vitrimers achieve both molecular variety and desirable $T_\mathrm{g}$ within 2 K range around the target, which make them ideal candidates for sustainable polymers for different applications. To validate our framework in experiments, we synthesize and characterize a novel vitrimer designed by the model. This vitrimer is proposed by optimizing acid-specific dimensions of latent vector while fixing epoxide as DGEBA. Driven by chemical intuition, we then slightly modify the proposed acids to a thermodynamically favorable derivative. The experimentally measured $T_\mathrm{g}$ (311 to 317 K) agrees well with the design target (323 K), which validates effectiveness of the VAE model. With slight adjustment to the VAE architecture and established computational methods to generate property data, the proposed framework can be extended to a wide range of properties (for example, Young's modulus, thermal conductivity, and topology freezing temperature) and other types of polymers. The complete workflow of computational design and experimental validation opens an opportunity for polymer scientists to achieve high-fidelity inverse design of multi-component polymeric materials with desirable properties.

\section*{Methods}

Details of MD simulations (Section \ref{supp_md}), GP calibration model (Section \ref{supp_gp}), hierarchical representation of molecules (Section \ref{supp_repre}), VAE architecture and training protocols (Section \ref{supp_vae}), VAE performance (Section \ref{supp_performance}), exploration of the latent space (Section \ref{supp_explore}), Bayesian optimization (Section \ref{supp_bo}), estimate of computational efficiency (Section \ref{supp_efficiency}) and experiments (Section \ref{supp_experiment}) are provided in Supplementary Information.

\section*{Data and code availability}

The code and data used for the GP calibration model and code used for the VAE framework and Bayesian optimization have been uploaded to GitHub (\url{https://github.com/yiwenzheng98/VitrimerVAE}). The generated vitrimer data (molecules in SMILES, density-temperature profiles and calculated $T_\mathrm{g}$) will be made openly available at the time of publication.

\section*{Acknowledgements}

Y. Zheng and A. Vashisth would like to thank Microsoft Climate Research Initiative and University of Washington, Seattle for funding. The research design, GPU for machine learning, and molecular simulations used in this study were partially provided by the HYAK supercomputer system of University of Washington. P. Thakolkaran and S. Kumar acknowledge the research funding by the Dutch Research Council (NWO): OCENW.XS22.2.103.

\section*{Authors contributions}

\textbf{Y.Z.}: Methodology, Software, Validation, Data Curation, Visualization, Writing, Reviewing and Editing; \textbf{P.T.}: Methodology, Software, Visualization, Reviewing and Editing; \textbf{A.K.B.}: Methodology, Experiment, Reviewing and Editing; \textbf{J.A.S.}: Conceptualization, Methodology, Software, Reviewing and Editing; \textbf{Z.L.}: Software, Reviewing and Editing; \textbf{S.Z.}: Reviewing and Editing; \textbf{B.H.N.}: Conceptualization, Reviewing and Editing, Supervision; \textbf{S.K.}: Conceptualization, Methodology, Reviewing and Editing, Supervision; \textbf{A.V.}: Conceptualization, Methodology, Reviewing and Editing, Supervision.

\section*{Competing interests}

\textbf{J.A.S.}, \textbf{Z.L.}, \textbf{S.Z.} and \textbf{B.H.N.} are employees of Microsoft Corporation. \textbf{Y.Z.}, \textbf{P.T.}, \textbf{A.K.B.}, \textbf{S.K.} and \textbf{A.V.} declare no competing interests.

\bibliography{sample}

\makeatletter\@input{yy.tex}\makeatother

\end{document}


\flushbottom
\maketitle

\section{Molecular dynamics simulations} \label{supp_md}

\begin{figure}[t]
\centering
\includegraphics[width=\linewidth]{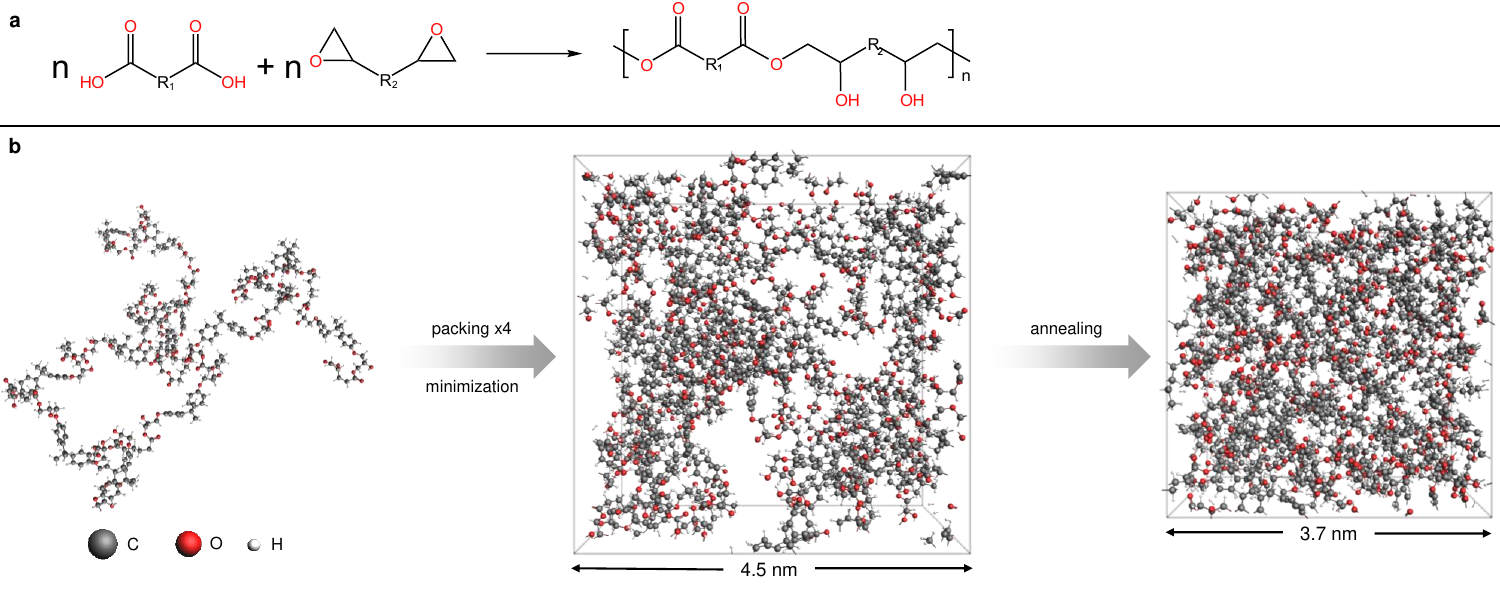}
\caption{\textbf{Molecular dynamics simulations to calculate $T_\mathrm{g}$ of vitrimers.} \textbf{(a)} A virtual vitrimer chain is made by connecting carboxylic acids and opened epoxides in an alternating manner. \textbf{(b)} Four chains of $\sim$ 1,000 atoms are placed in a simulation box. The system is annealed to remove local heterogeneities before production.}
\label{md}
\end{figure}

We perform MD simulations on 8,424 vitrimers sampled from the large dataset of one million to calculate their $T_\mathrm{g}$. The simulations are conducted by Large-scale Atomic/Molecular Massively Parallel Simulator (LAMMPS) \cite{thompson2022lammps} and Polymer Consistent Force Field (PCFF) \cite{sun1994ab} to describe the potential energy of atoms. PCFF has been applied to simulate the behavior of various polymeric systems including vitrimers \cite{sun2019molecular, park2020enhanced}. We build the polymer chains in an alternating copolymer manner. For each vitrimer composition (i.e., one carboxylic acid molecule and one epoxide molecule), we connect the acid molecule and opened epoxide molecule alternately to form one vitrimer chain of around one thousand atoms. The reaction scheme is shown in Supplementary Figure \ref{md}a. Four of these chains are placed in a cubic simulation box with a density of $0.5~\mathrm{g}/\mathrm{cm}^3$ (Supplementary Figure \ref{md}b). The connection and placement of atoms in the simulation box are done by Enhanced Monte Carlo package \cite{in2003temperature}, which creates input structures for MD simulations using the Monte Carlo method with energetically favored orientations.

The initial configuration is minimized using the conjugate gradient method and annealed to remove local heterogeneities. Specifically, the minimized structure is first relaxed under NVT ensemble (300 K) for 50 ps and under NPT ensemble (300 K and 1 atm) for 100 ps. We further heat the virtual specimen from 300 K to 800 K under NPT (1 atm) in 500 ps. The dimension of the simulation box is reduced and the density is taken to a realistic level after annealing. Two snapshots of the virtual specimens of an example vitrimer (adipic acid and bisphenol A diglycidyl ether) before and after annealing are presented in Supplementary Figure \ref{md}b. We hold the annealed system at 800 K for an additional 50 ps to obtain five independent specimens separated by 10 ps, which is proved to be sufficient to eliminate the effect of initial structures and ensure better statistics \cite{alzate2018uncertainties}. For production period, each of these specimens is cooled from 800 K to 100 K in a 10 K step. Each cooling step takes 25 ps under NPT followed by a 25-ps holding at constant temperature, during which the density is calculated as the average from 25 frames. We fit a bilinear regression to the density-temperature profile from 800 K to 100 K and the intersection point is defined as the $T_\mathrm{g}^\mathrm{MD}$ (Figure \ref{data}a in the manuscript). All $T_\mathrm{g}^\mathrm{MD}$ results from five virtual specimens are averaged to reduce the uncertainty due to the stochastic nature of MD simulations. The distributions of mean MD-calculated $T_\mathrm{g}^\mathrm{MD}$ and coefficient of variation (i.e., mean divided by standard deviation) from five replicates of 8,424 vitrimers are presented in Supplementary Figure \ref{tghist}.

\begin{figure}[!ht]
\centering
\includegraphics[width=\linewidth]{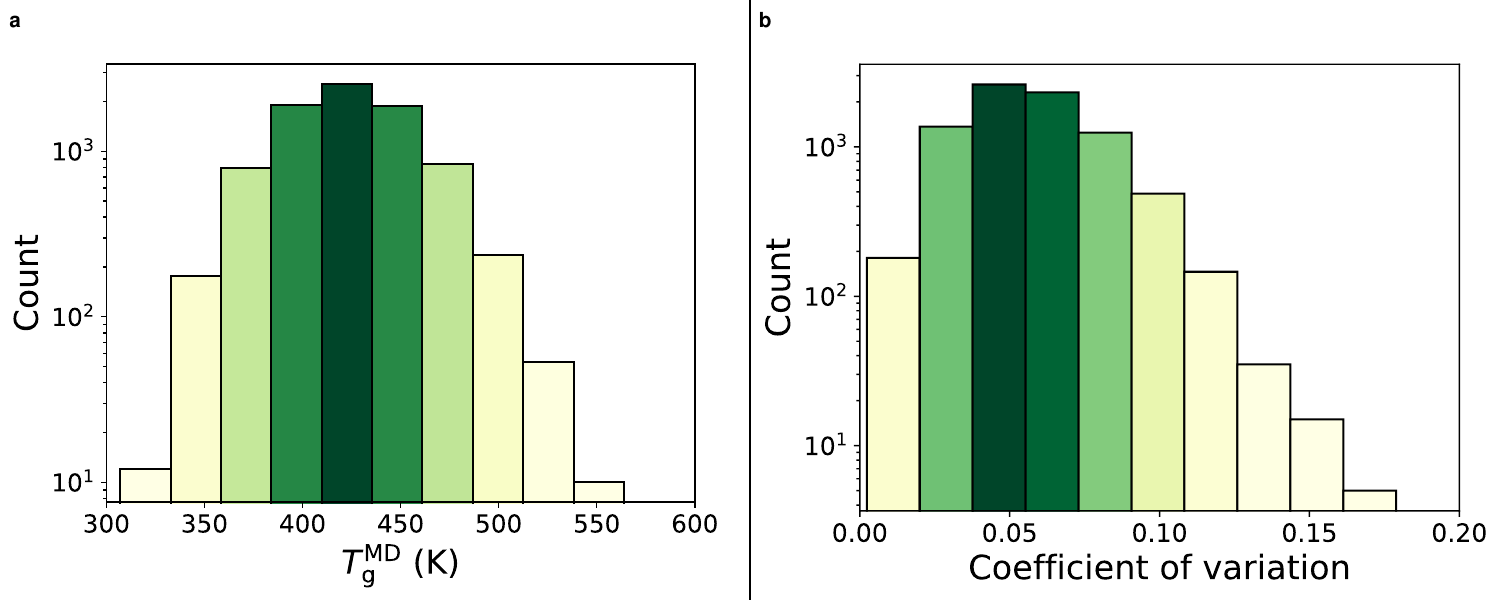}
\caption{Distributions of \textbf{(a)} mean $T_\mathrm{g}$ and \textbf{(b)} coefficient of variation in $T_\mathrm{g}$ from five replicate MD simulations of 8,424 vitrimers.}
\label{tghist}
\end{figure}

\section{Gaussian process model for calibration of $T_\mathrm{g}$} \label{supp_gp}

\begin{figure}[t]
\centering
\includegraphics[width=\linewidth]{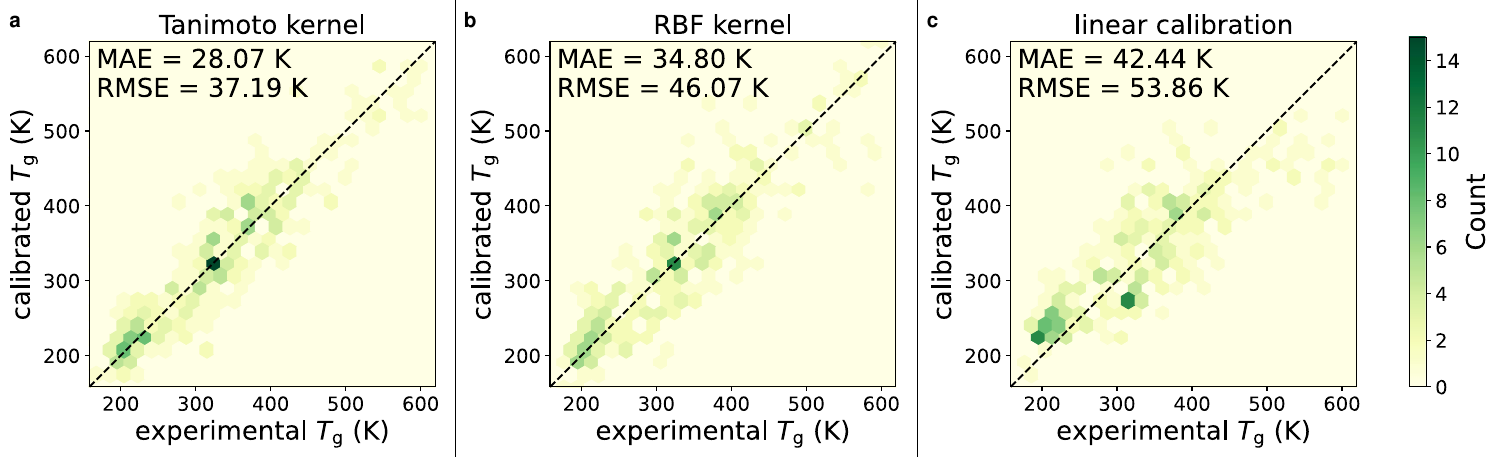}
\caption{\textbf{Calibrated $T_\mathrm{g}$ vs. experimental $T_\mathrm{g}$ by (a) GP with Tanimoto kernel, (b) GP with RBF kernel and (c) linear regression.}}
\label{calibration}
\end{figure}

To compensate for the overestimation in $T_\mathrm{g}^\mathrm{MD}$ due to the larger cooling rate compared with experiments, a calibration procedure is required. Afzal et al. \cite{afzal2020high} calibrate $T_\mathrm{g}^\mathrm{MD}$ against experimental $T_\mathrm{g}$ by linear regression. However, we find it insufficient due to the effect of increased uncertainty associated with smaller systems in our simulations (Supplementary Figure \ref{calibration}c). Instead, we calibrate $T_\mathrm{g}^\mathrm{MD}$ from MD simulations against available experimental data in literature using a Gaussian process (GP) regression model. 2048-bit extended-connectivity fingerprints (ECFPs) \cite{rogers2010extended} with radius 3 are used as input to the GP model to represent the repeating units of polymers. For vitrimers (i.e., combinations of carboxylic acids and epoxides), we obtain the repeating units by the reaction scheme depicted in Supplementary Figure \ref{md}a with $n = 1$. For the GP model, we employ the Tanimoto kernel \cite{ralaivola2005graph, thawani2020photoswitch}, which relies on the Tanimoto similarity measure extensively used in cheminformatics. Unlike the traditional kernels (e.g., the radial basis function kernel used by Jinich et al. \cite{jinich2019mixed}) which are more suitable for continuous spaces (Supplementary Figure \ref{calibration}b), Tanimoto kernel allows for swift and accurate comparison between molecular fingerprints as bit vectors. The Tanimoto kernel in the GP model is defined as follows:
\begin{align}
    k(\boldsymbol{x}, \boldsymbol{x}') = \frac{\sigma \langle \boldsymbol{x}, \boldsymbol{x}' \rangle}{\|\boldsymbol{x}\|^2 + \|\boldsymbol{x}'\|^2 - \langle \boldsymbol{x}, \boldsymbol{x}' \rangle}, 
\end{align}
where $\boldsymbol{x}, \boldsymbol{x}'$ are fingerprints of two molecules, $\sigma$ is the variance of the kernel, $\langle \boldsymbol{x}, \boldsymbol{x}' \rangle$ is the inner product and $\|\boldsymbol{x}\|^2$ is the square of norm. Calibrated and experimental $T_\mathrm{g}$  from leave-one-out cross validation (LOOCV) with different calibration methods are compared in Supplementary Figure \ref{calibration} and GP calibration with Tanimoto kernel achieves the best accuracy with a mean absolute error of 28.07 K. We proceed to calibrate $T_\mathrm{g}^\mathrm{MD}$ of all 8,424 vitrimers by the trained GP model. Ten vitrimers with highest and lowest calibrated $T_\mathrm{g}$ are presented in Supplementary Figure \ref{tg}.

\begin{figure}[!ht]
\centering
\includegraphics[width=\linewidth]{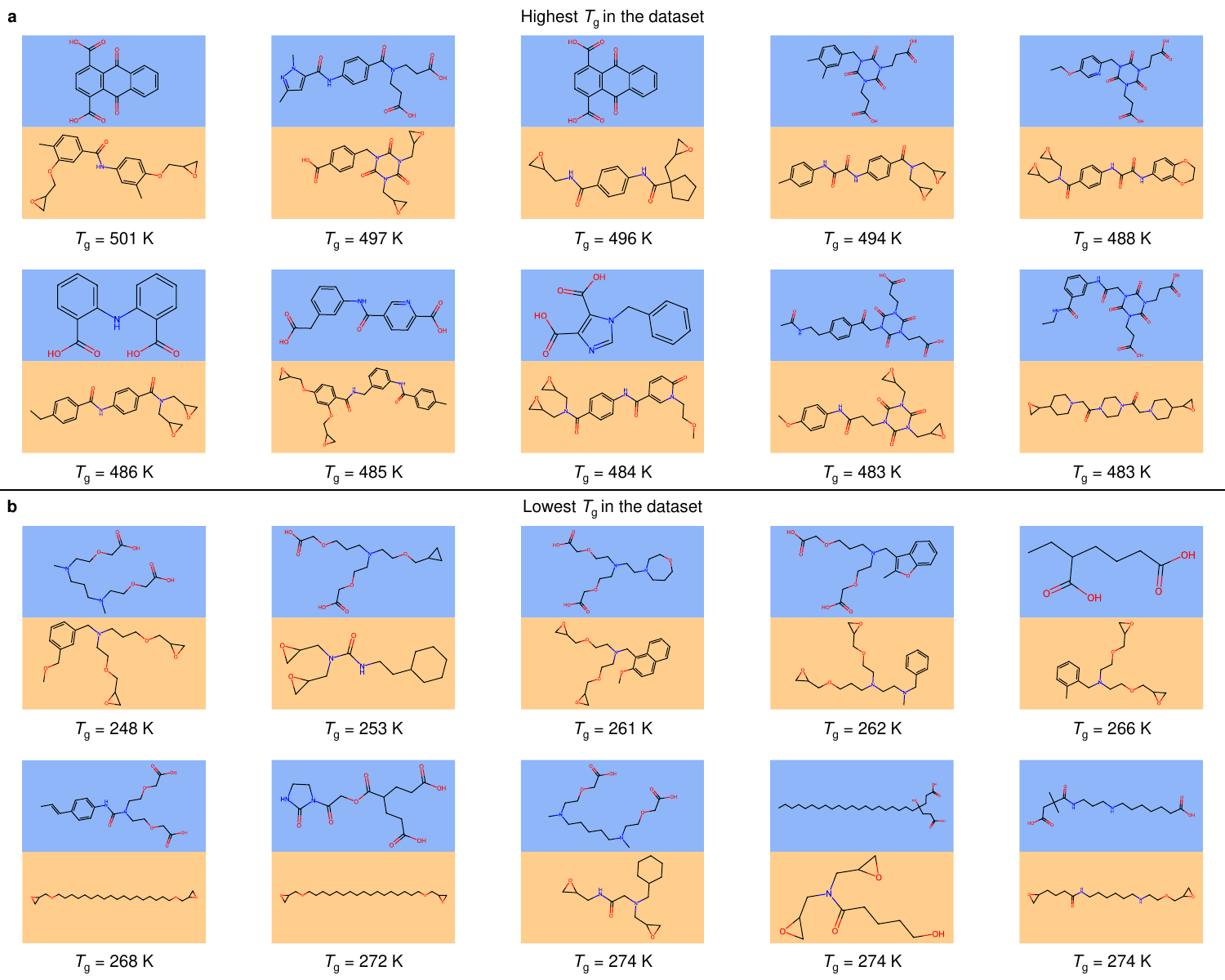}
\caption{\textbf{Ten example vitrimers with highest and lowest $T_\mathrm{g}$ in the dataset.}}
\label{tg}
\end{figure}

\section{Machine learning framework} \label{supp_ml}

\subsection{Hierarchical representation of molecules} \label{supp_repre}

\begin{figure}[t]
\centering
\includegraphics[width=.7\linewidth]{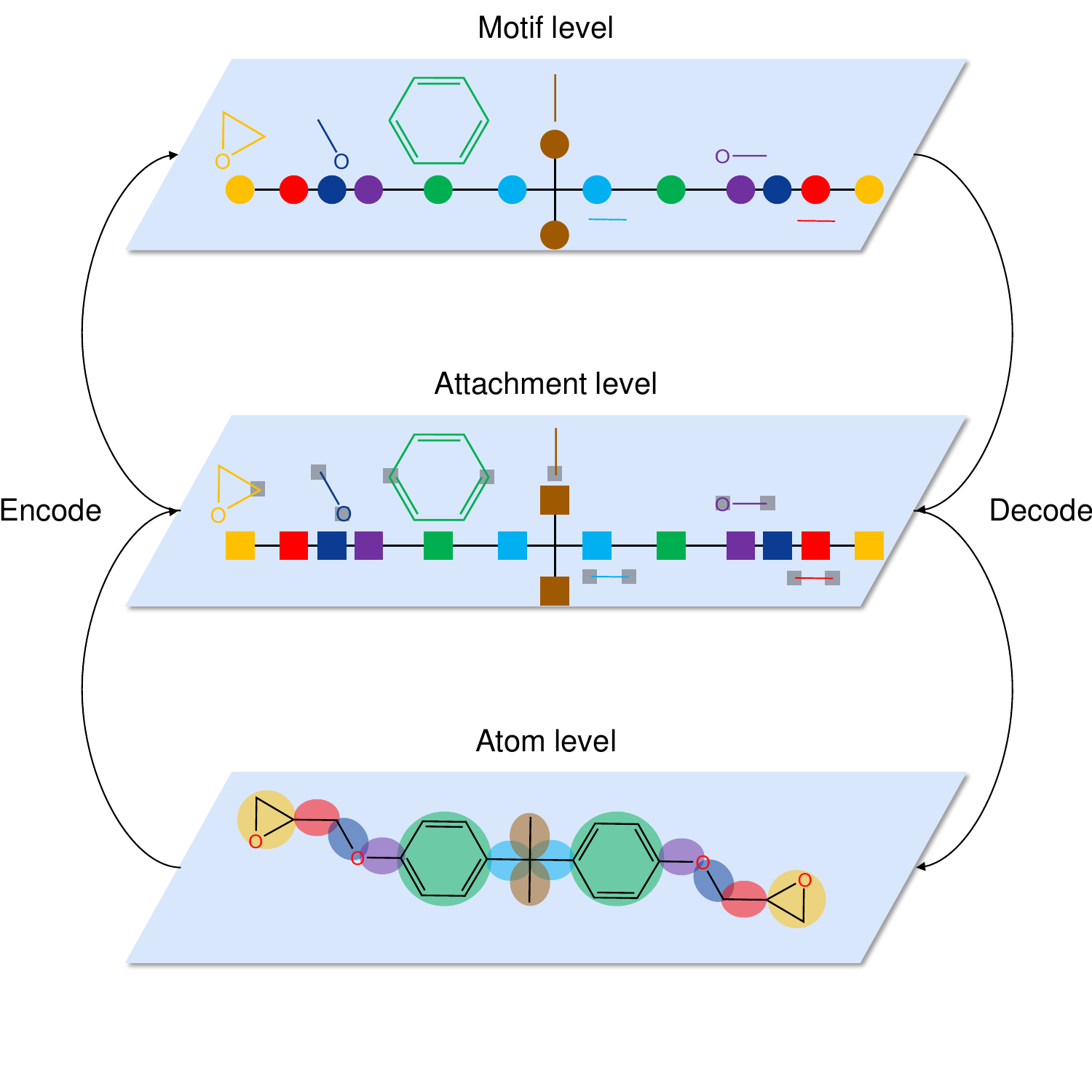}
\caption{\textbf{Schematic illustration of the three-level hierarchical graph representation.} The shared atoms between motifs are highlighted in grey blocks on the attachment level.}
\label{hierarchical}
\end{figure}

We follow the hierarchical representation of molecules proposed by Jin et al. \cite{jin2020hierarchical}. A molecule $\mathcal{G} = (\mathcal{V}, \mathcal{E})$ is preprocessed by decomposing it into $n$ motifs (subgraphs) $\mathcal{M}_1, ..., \mathcal{M}_n$ using the following procedures. First, we identify all bonds between atom pair $u$ and $v$ with more than two connections and either $u$ or $v$ belongs to a ring. We then break all these bonds and convert $\mathcal{G}$ into a series of detached graphs. For each detached graph, if it appears more than a threshold frequency $f$ during decomposition of all molecules in the training set, it is selected as a motif. Otherwise it is further decomposed into single rings and bonds (i.e., one bond with two atoms) which are selected as motifs. In this way, chemical validity is preserved and the union of all motifs covers the entire graph. We preprocess all molecules in the training set by the same procedures and a motif vocabulary $V_\mathcal{M}$ is obtained. We empirically find that the VAE model works better when the motifs are constrained to single rings and bonds, i.e., $f \rightarrow \infty$. This can be attributed to the fact that all molecules in the dataset are relatively small and limited to a maximum molecular weight of 500 g/mol. Larger motifs with more than one bonds can be used to accommodate larger molecules if necessary. Two vocabularies with sizes of 124 and 104 motifs are constructed for acid and epoxide molecules, respectively.

We further represent $\mathcal{G}$ as a combination of hierarchical graphs at three levels (see Supplementary Figure \ref{hierarchical} for a schematic illustration of the three-level hierarchical representation). The motif level $\mathcal{G}_\mathcal{M}$ captures how motifs $\mathcal{M}_1, ..., \mathcal{M}_n$ are connected, i.e., $\mathcal{G}_\mathcal{M} = (\mathcal{V}_\mathcal{M}, \mathcal{E}_\mathcal{M})$ with motifs as nodes and bonds connecting motifs as edges. In addition, the attachment level $\mathcal{G}_\mathcal{A}$ represents connection between motifs through shared atoms. Each node $\mathcal{A}_i = (\mathcal{M}_i, \{v_j\})$ at this level defines a connection site of $\mathcal{M}_i$ with $\{v_j\}$ as all possible atoms shared by $\mathcal{S}_i$ and its neighbors. Since the possible connection sites of $\mathcal{M}_i$ are finite, a vocabulary of attachment nodes $V_\mathcal{A}(\mathcal{M}_i)$ that depends on each  motif $\mathcal{M}_i$ is constructed. Finally, the atom level $\mathcal{G}$ encodes the graph at the atomic level. Each node is an atom and each edge is a bond of the molecule.
This representation scheme captures necessary information of molecules at three levels with different resolutions and ensures chemical validity. As a result, the associative encoder and decoder can achieve accurate reconstruction and efficient generation of valid vitrimers in the VAE framework.

\subsection{VAE architecture and training protocols} \label{supp_vae}

We adopt the hierarchical encoder and decoder associated with the hierarchical graph representation \cite{jin2020hierarchical}. We use PyTorch \cite{paszke2019pytorch} to implement and train the model, and RDKit \cite{rdkit} for cheminformatics operations.

\subsubsection{Encoder}

The hierarchical encoder encodes a molecule $\mathcal{G}$ (which can be either acid or epoxide; superscripts omitted for simplicity) from the finer level to the coarser level. The encoder contains three message passing networks (MPNs) as detailed by Jin et al. \cite{jin2020hierarchical}. At the atom level, each node is an atom $v$ and each edge is a bond $e_{uv}$ between atoms $u$ and $v$. The node features consist of atomic charge and atom type and the edge feature is bond type. All atom and bond types are represented as one-hot encodings. The node and edge features are first converted into embedding vectors before passing to the MPN:
\begin{equation}
    \{\boldsymbol{h}_v\} = \mathrm{MPN} \left(\mathcal{G}, \{E(v)\}, \{E(e_{uv})\}\right),
\end{equation}
where $\{\boldsymbol{h}_v\}$ is the atom-level encoding for each atom $v$ and $E(\boldsymbol{\cdot})$ denotes embedding vector of $(\boldsymbol{\cdot})$. At the attachment level, the node feature is a concatenation of its embedding $E(\mathcal{A}_i)$ and the sum of all atom-level encodings of its constituent atoms:
\begin{equation}
    \boldsymbol{f}_{\mathcal{A}_i} = \mathrm{MLP} \left(E(\mathcal{A}_i) \oplus \sum_{v \in \mathcal{M}_i} \boldsymbol{h}_v\right).
\end{equation}
The edge feature between attachment nodes $\mathcal{A}_i$ and $\mathcal{A}_j$ is an embedding of a parameter $x_{ij}$ denoting their parent-child relation based on depth-first search:
\begin{equation}
    x_{ij} = 
    \begin{cases}
        0,& \text{if } \mathcal{A}_j \text{ is the parent}; \\
        k,& \text{if } \mathcal{A}_i \text{ is the } k \text{-th child of } \mathcal{A}_j.
    \end{cases}
\end{equation}
The attachment-level encodings are calculated as
\begin{equation}
    \{\boldsymbol{h}_{\mathcal{A}_i}\} = \mathrm{MPN} \left(\mathcal{G}_\mathcal{A}, \{\boldsymbol{f}_{\mathcal{A}_i}\}, \{E(x_{ij})\}\right).
\end{equation}
The encoding at the motif level is similar to attachment level with nodes as motifs $\mathcal{M}_i$, i.e., 
\begin{equation}
    \boldsymbol{f}_{\mathcal{M}_i} = \mathrm{MLP} \left(E(\mathcal{M}_i) \oplus \boldsymbol{h}_{\mathcal{A}_i}\right).
\end{equation}
and
\begin{equation}
    \{\boldsymbol{h}_{\mathcal{M}_i}\} = \mathrm{MPN} \left(\mathcal{G}_\mathcal{M}, \{\boldsymbol{f}_{\mathcal{M}_i}\}, \{E(x_{ij})\}\right).
\end{equation}
Two linear neural networks are used to output mean vector $\boldsymbol{\mu}$ and log variance vector $\log \boldsymbol{\sigma}^2$ from the encoding of the root motif $\boldsymbol{h}_{\mathcal{M}_1}$. The root motif is the first motif to be generated during decoding.

\subsubsection{Decoder}

The decoder attempts to iteratively build the hierarchical graph based on latent vector $\boldsymbol{z}$ of the original molecule. Here for simplicity we refer to $\boldsymbol{z}$ as the concatenation of acid-specific dimensions (or epoxide-specific dimensions) with shared dimensions (see Equation \ref{partial} in the manuscript). At $t$-th step of generation, we denote $\mathcal{M}_k$ as the motif whose neighbor will be generated in the next step. Motifs are generated in a depth-first order. We use the same hierarchical MPN architecture to encode all the motifs and atoms in the partially generated graph and obtain motif encodings $\boldsymbol{h}_{\mathcal{M}_k}$ and atom encodings $\boldsymbol{h}_v$ for each existing motif and atom. At the motif level, the next motif $\mathcal{M}_t$ to be connected to $\mathcal{M}_k$ is predicted based on the entire motif vocabulary $V_\mathcal{M}$:
\begin{equation}
    \boldsymbol{p}_{\mathcal{M}_t} = \mathrm{softmax}\left(\mathrm{MLP}\left(\boldsymbol{h}_{\mathcal{M}_k}\oplus \boldsymbol{z}\right)\right).
\end{equation}
At the attachment level, we predict which attachment $\mathcal{A}_t$ of motif $\mathcal{M}_t$ is used, which is classified over the attachment vocabulary $V_\mathcal{A}(\mathcal{M}_t)$:
\begin{equation}
    \boldsymbol{p}_{\mathcal{A}_t} = \mathrm{softmax}\left(\mathrm{MLP}\left(\boldsymbol{h}_{\mathcal{M}_k}\oplus \boldsymbol{z}\right)\right).
\end{equation}
At the atom level, the detailed atomic attachment configuration $(u, v)$ is decided, where $u$ and $v$ are atoms from $\mathcal{A}_k$ and $\mathcal{A}_t$, respectively. The probability of a certain attachment configuration $(u, v)_t$ is calculated as
\begin{equation}
    \boldsymbol{p}_{(u,v)_t} = \mathrm{softmax} \left(\mathrm{MLP}\left(\boldsymbol{h}_u \oplus \boldsymbol{h}_v\right) \cdot \boldsymbol{z}\right).
\end{equation}
An additional MLP is used to predict the probability of backtracing, i.e., there are no new neighbors to be generated for $\mathcal{M}_k$:
\begin{equation}
    \boldsymbol{p}_{\mathrm{b}_t} = \mathrm{softmax}\left(\mathrm{MLP}\left(\boldsymbol{h}_{\mathcal{M}_k}\oplus \boldsymbol{z}\right)\right).
\end{equation}

\subsubsection{Training}

\begin{table}[t]
    \centering
    \begin{tabular}{c c c c}
        \Xhline{2\arrayrulewidth}
        \textbf{Hyperparameters} & \textbf{Encoder} $\mathcal{Q}_\phi$ & \textbf{Decoder} $\mathcal{P}_\theta$ & \textbf{Property predictor} $\mathcal{F}_\omega$ \\
        \hline
        Input dimensions & none & $d_\mathrm{a}=d_\mathrm{e}=112$ & $d=128$ \\
        Embedding dimensions & 250 & 250 & none \\
        Hidden dimensions & 250 & 250 & 64 \\
        Output dimensions & $d_\mathrm{a}=d_\mathrm{e}=112$ & none & 1 \\
        \Xhline{2\arrayrulewidth}
    \end{tabular}
    \caption{\textbf{Hyperparameters of the network architecture.}}
    \label{hyper1}
\end{table}

\begin{table}[]
    \centering
    \begin{tabular}{c c c}
        \Xhline{2\arrayrulewidth}
        \textbf{Hyperparameters} & \textbf{Step one} & \textbf{Step two} \\
        \hline
        $\lambda_\mathrm{KL}$ & 0.005 & 0.005 \\
        Batch size & 32 & 32 \\
        Optimizer & Adam \cite{kingma2014adam} & Adam \cite{kingma2014adam} \\
        Learning rate & constant 0.001 & $0.001 \times 0.9^{i-1}$ at epoch $i$\\
        Number of epochs & 10 & 50 \\
        \Xhline{2\arrayrulewidth}
    \end{tabular}
    \caption{\textbf{Hyperparameters of the training protocols.}}
    \label{hyper2}
\end{table}

The one million vitrimer dataset is divided into two subsets: 999,000 vitrimers without $T_\mathrm{g}$ as unlabeled training set $\mathcal{D} = \{({\mathcal{G}^\mathrm{a}}^{(i)}, {\mathcal{G}^\mathrm{e}}^{(i)}): i = 1, ..., 999000\}$ and 1,000 vitrimers with $T_\mathrm{g}$ as test set $\mathcal{D}_\mathrm{test} = \{({\mathcal{G}^\mathrm{a}}^{(i)}, {\mathcal{G}^\mathrm{e}}^{(i)}, T_\mathrm{g}^{(i)}): i = 1, ..., 1000\}$. We randomly sample a subset of 8,424 vitrimers from the whole dataset and calculate their $T_\mathrm{g}$ by MD simulations and GP calibration. A subset of 7,424 vitrimers is denoted as the labeled training set $\mathcal{D}_\mathrm{prop} = \{({\mathcal{G}^\mathrm{a}}^{(i)}, {\mathcal{G}^\mathrm{e}}^{(i)}, T_\mathrm{g}^{(i)}): i = 1, ..., 7424\}$ and the rest 1,000 vitrimers constitute the test set $\mathcal{D}_\mathrm{test}$. $\mathcal{D}$ and $\mathcal{D}_\mathrm{prop}$ are used to train the VAE on a two-step basis (see Equation \ref{loss1} and Equation \ref{loss2} in the manuscript) and $\mathcal{D}_\mathrm{test}$ allows for evaluation of the reconstruction and property prediction capabilities on unseen data. The network dimensions and hyperparameters of the VAE framework are presented in Supplementary Tables \ref{hyper1} and \ref{hyper2}, respectively. Note that we do not differentiate between the acid and epoxide encoders (decoders) because the network architectures are identical.

\subsection{VAE performance} \label{supp_performance}

\begin{figure}[!ht]
\centering
\includegraphics[width=\linewidth]{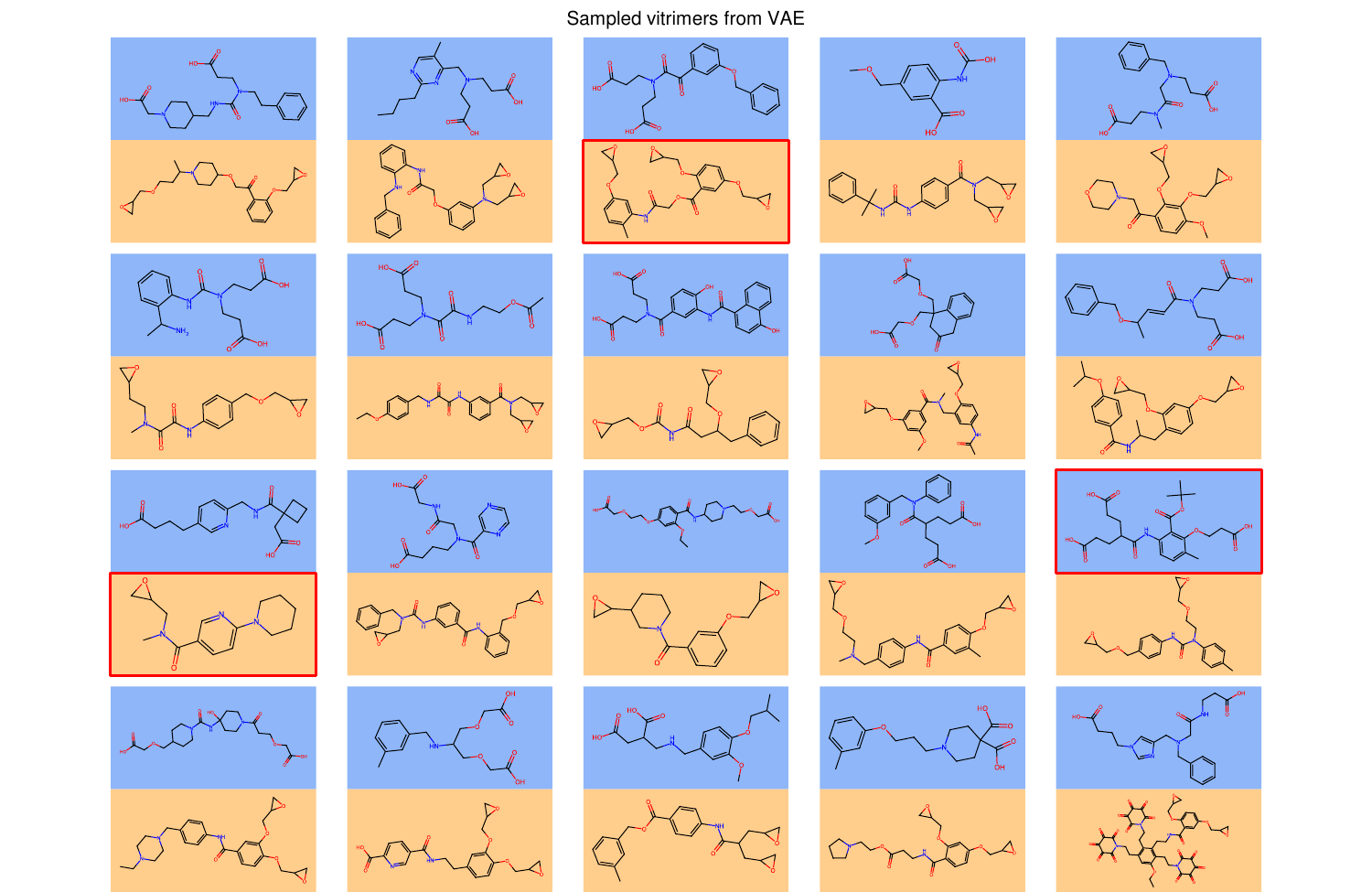}
\caption{\textbf{Examples of vitrimers sampled from the latent space according to standard Gaussian distribution.} Invalid vitrimer components are highlighted in red boxes.}
\label{sample}
\end{figure}

The metrics of the VAE model before and after joint training with the small dataset $\mathcal{D}_\mathrm{prop}$ are presented in Supplementary Table \ref{metrics}. The improved metrics show that the model is not biased to $\mathcal{D}_\mathrm{prop}$ and is able to accommodate and generate a wide range of vitrimers. Examples of ten vitrimers from the test set and their reconstructions are shown in Supplementary Figure \ref{reconstruct}. Two out of ten vitrimers are not successfully reconstructed due to mismatch of one of the components. Supplementary Figure \ref{sample} presents examples of 20 vitrimers sampled from the latent space based on standard Gaussian distribution. The three invalid vitrimers have chemically valid acids or epoxides which are not bifunctional. The predicted $T_\mathrm{g}$ by the property predictor is compared with calibrated $T_\mathrm{g}$ for 1,000 vitrimers in the test set, as shown in Supplementary Figure \ref{prediction}. A low MAE of 13.53 K indicates high accuracy of the trained property predictor. The distributions of latent vectors encoded from the labeled training set $\mathcal{D}_\mathrm{prop}$ and test set $\mathcal{D}_\mathrm{test}$ after joint training are shown in Supplementary Figure \ref{latent}a,b. Compare with the distributions before joint training (Supplementary Figure \ref{latent}c,d), the gradient in $T_\mathrm{g}$ is much more recognizable, proving the effect of latent space organization by joint training the VAE with property predictor.

\begin{table}[!ht]
    \centering
    \begin{tabular}{c c c}
        \Xhline{2\arrayrulewidth}
        \textbf{Metrics} & \textbf{Before joint training} & \textbf{After joint training} \\
        \hline
        Reconstruction accuracy & 81.8\% & 85.4\% \\
        Sample validity & 82.8\% & 85.2\% \\
        Sample novelty & 100.0\% & 100.0\% \\
        Sample uniqueness & 100.0\% & 100.0\% \\
        \Xhline{2\arrayrulewidth}
    \end{tabular}
    \caption{\textbf{Metrics of the VAE before and after joint training.}}
    \label{metrics}
\end{table}

\begin{figure}[!ht]
\centering
\includegraphics[width=\linewidth]{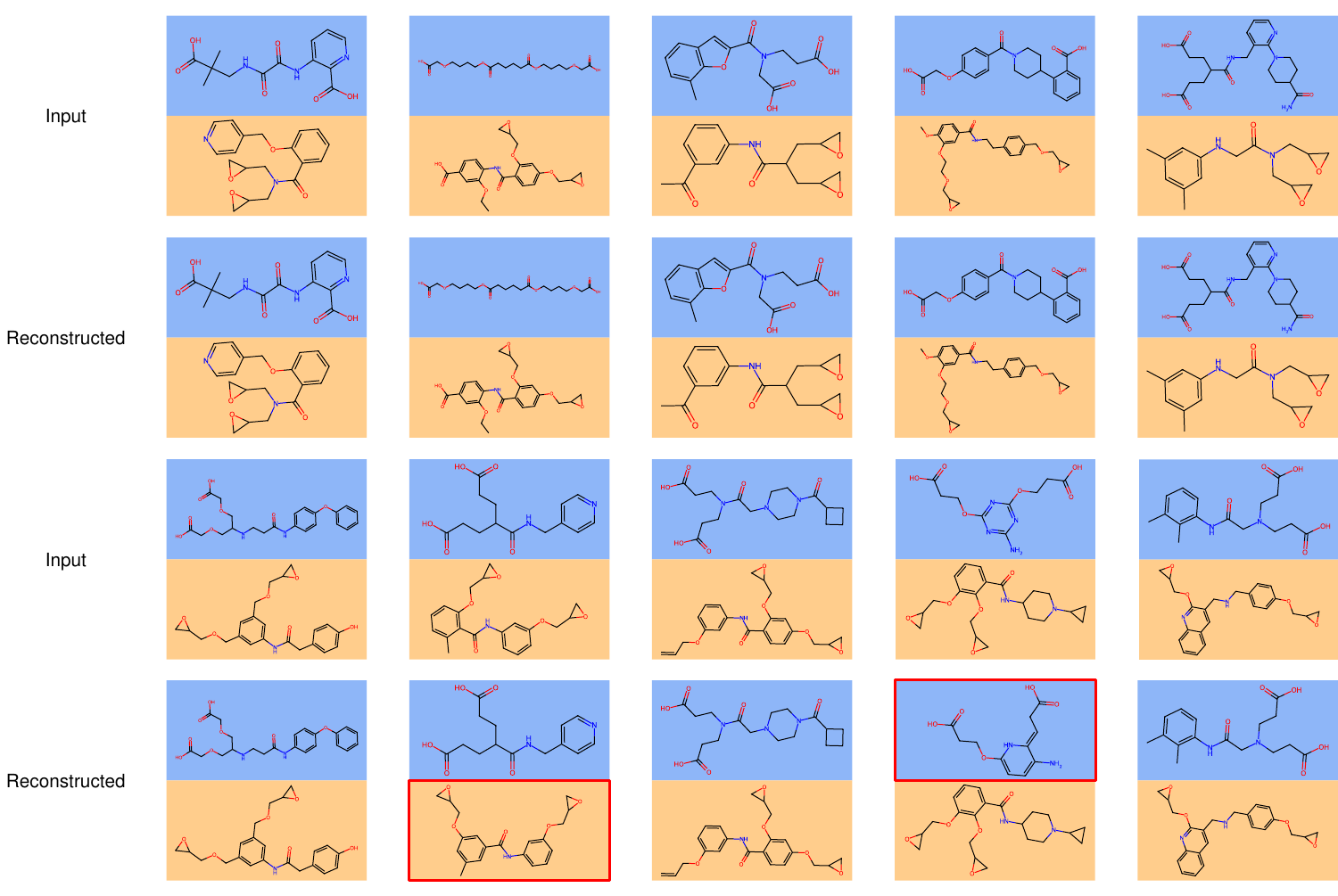}
\caption{\textbf{Examples of input and reconstructed vitrimers from the test set.} Unsuccessful reconstructions are highlighted in red boxes.}
\label{reconstruct}
\end{figure}

\begin{figure}[!ht]
\centering
\includegraphics[width=.5\linewidth]{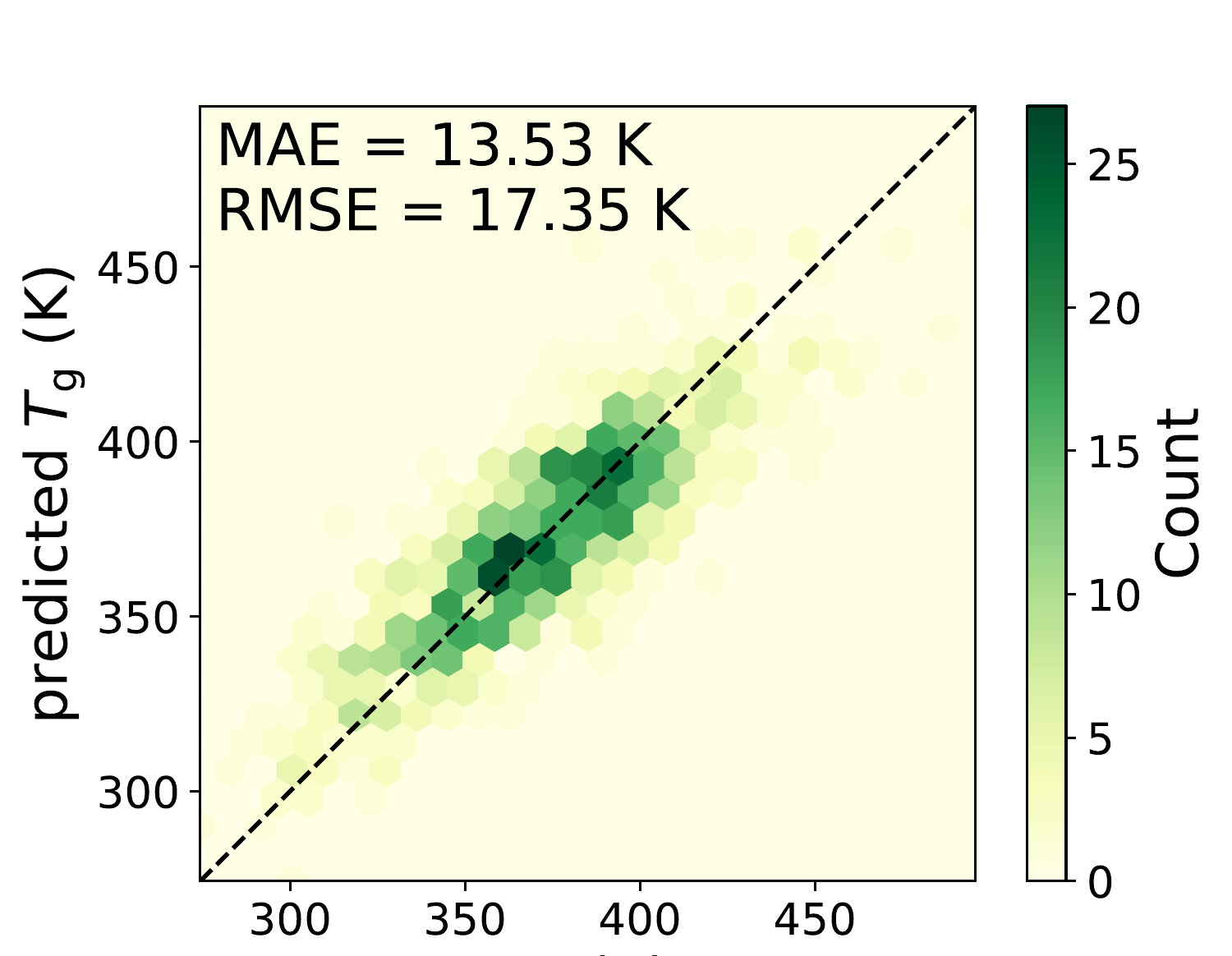}
\caption{\textbf{Predicted $T_\mathrm{g}$ by the property predictor vs. true $T_\mathrm{g}$.}}
\label{prediction}
\end{figure}

\begin{figure}[!ht]
\centering
\includegraphics[width=\linewidth]{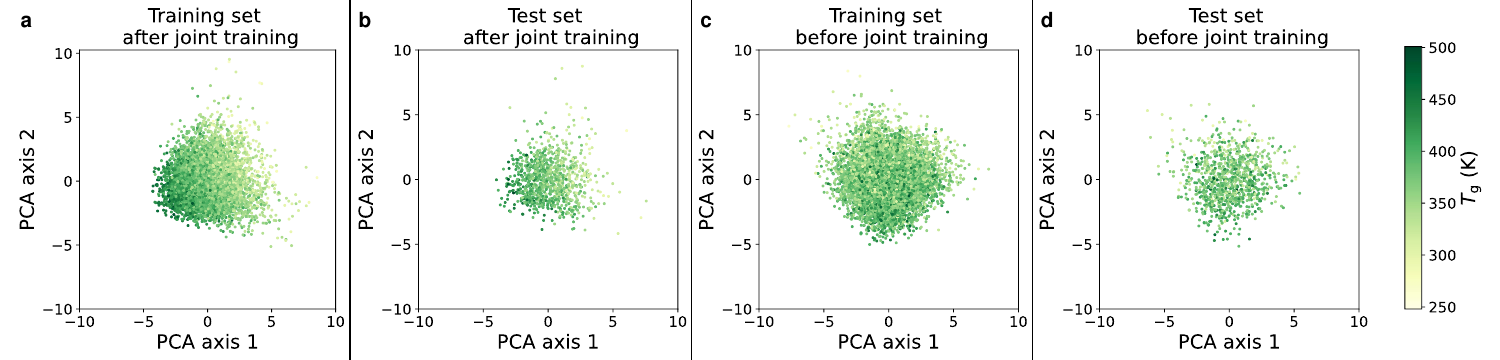}
\caption{\textbf{Distributions of latent vectors of (a) training set after joint training, (b) test set after joint training, (c) training set before joint training and (d) test set before joint training.} Latent vectors of higher dimensions are projected into two principal axes for visualization using principal component analysis (PCA).}
\label{latent}
\end{figure}

\subsection{Exploration of latent space} \label{supp_explore}

\begin{figure}[!ht]
\centering
\includegraphics[width=\linewidth]{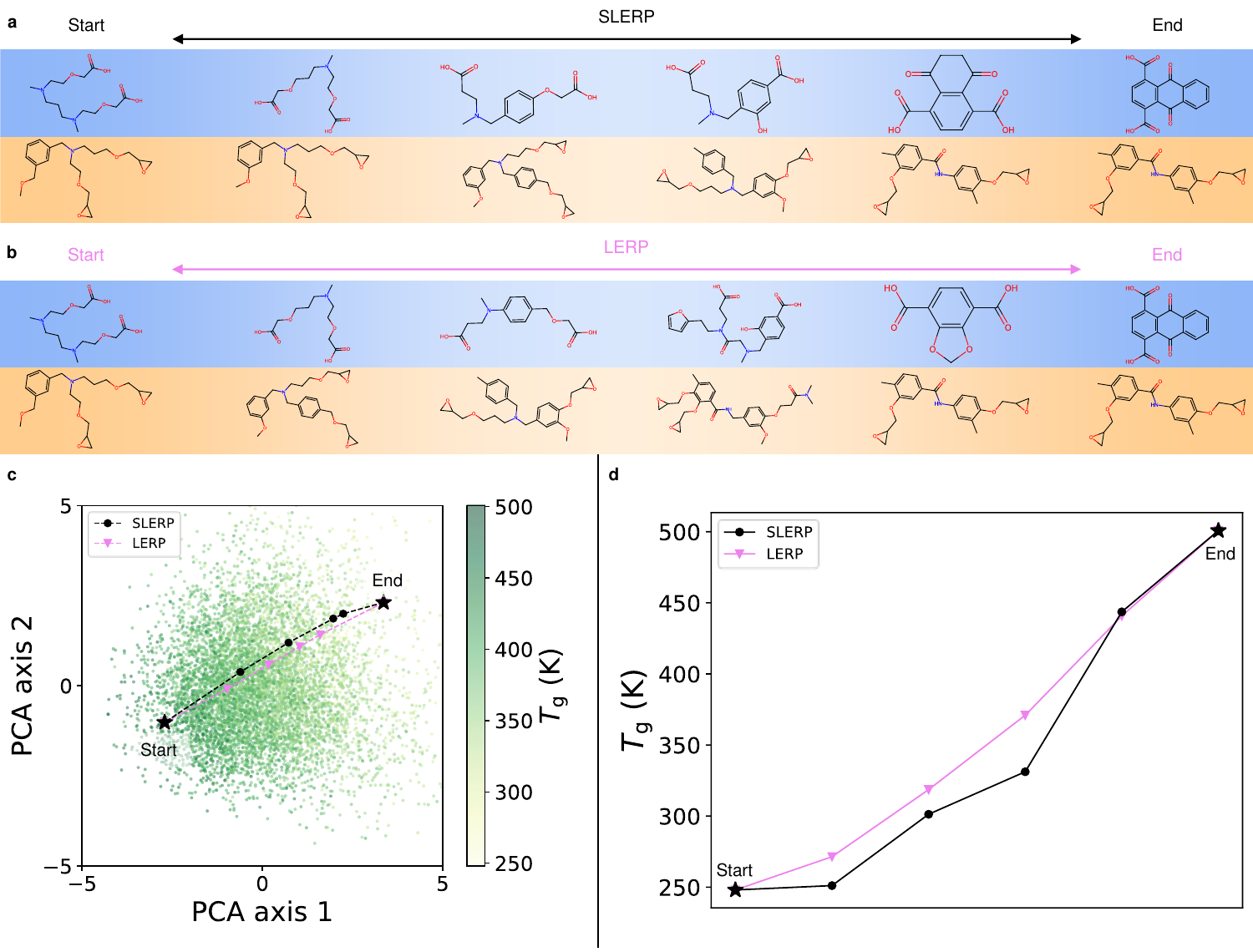}
\caption{\textbf{Comparison between spherical interpolation (SLERP) and linear interpolation (LERP).} \textbf{(a)}\textbf{(b)} Vitrimers discovered by \textbf{(a)} SLERP and \textbf{(b)} LERP in the latent space. \textbf{(c)} Interpolation paths in latent space visualized by PCA. \textbf{(d)} $T_\mathrm{g}$ of discovered vitrimers. All presented $T_\mathrm{g}$ values of are validated by MD simulations and GP calibration.}
\label{inter}
\end{figure}

Starting with the latent vector of a know vitrimer $\boldsymbol{z_0}$ as origin, we can find novel vitrimers by adding noise, i.e.,
\begin{equation}
    \boldsymbol{z} = \boldsymbol{z}_0 + \beta \cdot \boldsymbol{\epsilon},
\end{equation}
where $\beta$ determines the magnitude of the noise and $\boldsymbol{\epsilon} \sim \mathcal{N}(\boldsymbol{0}, \boldsymbol{I})$. The partial overlapping method enables us to explore the neighborhood along three axes: acid-specific (first $d_\mathrm{a}$ dimensions of $\boldsymbol{\epsilon}$ are non-zero), epoxide-specific (last $d_\mathrm{e}$ dimensions of $\boldsymbol{\epsilon}$ are non-zero) and both (all dimensions of $\boldsymbol{\epsilon}$ are non-zero).

We define a spherical interpolation (SLERP) path between two latent representations $\boldsymbol{z}_1$ and $\boldsymbol{z}_2$:
\begin{equation}
    \mathrm{SLERP}(\boldsymbol{z}_1, \boldsymbol{z}_2; \alpha) = \frac{\sin ((1-\alpha)\theta)}{\sin \theta} \boldsymbol{z}_1 + \frac{\sin (\alpha \theta)}{\sin \theta} \boldsymbol{z}_2,
\end{equation}
where $\alpha \in [0, 1]$ is the interpolation parameter and $\theta$ is the angle between $\boldsymbol{z}_1$ and $\boldsymbol{z}_2$. This is different from the linear interpolation (LERP) that has been used in previous works \cite{batra2020polymers, yao2021inverse}:
\begin{equation}
    \mathrm{LERP}(\boldsymbol{z}_1, \boldsymbol{z}_2; \alpha) = (1 - \alpha) \boldsymbol{z}_1 + \alpha \boldsymbol{z}_2.
\end{equation}
We choose SLERP over LERP since LERP operates on the presumption of a linear connection between points which disregards the inherent structure of the multi-dimensional Gaussian distribution. Moreover, LERP calculates the Euclidean distance between two points which may not be consistent with the similarity between vitrimers in the latent space. Therefore, points along the linear interpolation path might leap across regions within the latent space that share similar molecular structures which results in irregular or unnatural interpolations. On the other hand, SLERP acknowledges the hyperspherical structure of the latent space and pursues the shortest arc on the surface of a $d$-dimensional hypersphere, thereby minimizing unrealistic transitions along the path. Supplementary Figure \ref{inter} shows the decoded vitrimers along SLERP and LERP paths, their locations in the latent space and $T_\mathrm{g}$. Previous studies \cite{gomez2018automatic, white2016sampling} have addressed the effectiveness of SLERP, which is more suitable for traversing between two locations in the latent space created by VAEs.

\subsection{Bayesian optimization} \label{supp_bo}

We use the batch Bayesian optimization algorithm developed by Kusner et al. \cite{kusner2017grammar}. The workflow of Bayesian optimization is illustrated in Supplementary Figure \ref{boflow}. We start with 1,000 latent vectors $\boldsymbol{z}$ randomly sampled from standard Gaussian distribution and decode them into valid vitrimers. These vitrimers are then encoded into reconstructed latent vectors and their $T_\mathrm{g}$ is predicted by the property predictor. The predicted $T_\mathrm{g}$ is converted into the optimization objective, which is either $-T_\mathrm{g}$ (if the task is to find vitrimers with the maximum $T_\mathrm{g}$) or the squared error between $T_\mathrm{g}$ and the target $T_\mathrm{g}$ (if the task is to find vitrimers with a target $T_\mathrm{g}$). In each iteration, a sparse Gaussian process with 100 inducing points is trained as a surrogate model and 50 vitrimers are proposed by the expected improvement acquisition function. The valid vitrimers and their predicted $T_\mathrm{g}$ are added to the training set for next iteration. Ten independent runs each with 50 iterations are performed and all valid proposed vitrimers are collected. For each of the three design targets, we select 100 candidates with predicted $T_\mathrm{g}$ closest to the target that can be parameterized by PCFF, carry out MD simulations and calibrate MD-calculated $T_\mathrm{g}^\mathrm{MD}$. The distributions of $T_\mathrm{g}$ of the designed vitrimers and the dataset are presented in Supplementary Figure \ref{violin}. Ten best candidates with $T_\mathrm{g}$ closet to each target are presented in Supplementary Figure \ref{morebo}. Our framework succeeds in extrapolating beyond the training regime and achieves accurate inverse design within an error of 2 K.

To examine the stability of vitrimers presented in Supplementary Figure \ref{morebo}, we minimize them under reactive force field (ReaxFF). ReaxFF is a molecular dynamics simulation technique that captures the instantaneous interactions between atoms using the bond order concept \cite{van2001reaxff}. This strategy allows for a smooth transition between the non-bonded and bonded configurations and helps with the investigation of chemical reactions where bond formation and dissociation are involved. The proposed molecules remain intact during minimization using the CHON2017\_weak\_bb force field \cite{vashisth2018accelerated} and the minimized structures are presented in Supplementary Figure \ref{minimized}. We further calculate relevant molecular descriptors of these vitrimers, as shown in Supplementary Figure \ref{descriptor}. Density is extracted from MD simulations at 300 K. All other descriptors are calculated by the Mordred package \cite{moriwaki2018mordred}. Each value is calculated from the repeating units (Supplementary Figure \ref{md}a with $n = 1$) and averaged over ten proposed vitrimers.

We compare the $T_\mathrm{g}$ of the proposed vitrimers with commonly used polymers in Supplementary Figure \ref{tgbarchart}. The vitrimers from inverse design cover a wide range of $T_\mathrm{g}$ from around 250 K to 550 K. With further tuning of the design target, our framework has the potential to discover vitrimers with any $T_\mathrm{g}$ within the range and greatly enhances the applicability of vitrimers at various temperatures.

\begin{figure}[!ht]
\centering
\includegraphics[width=\linewidth]{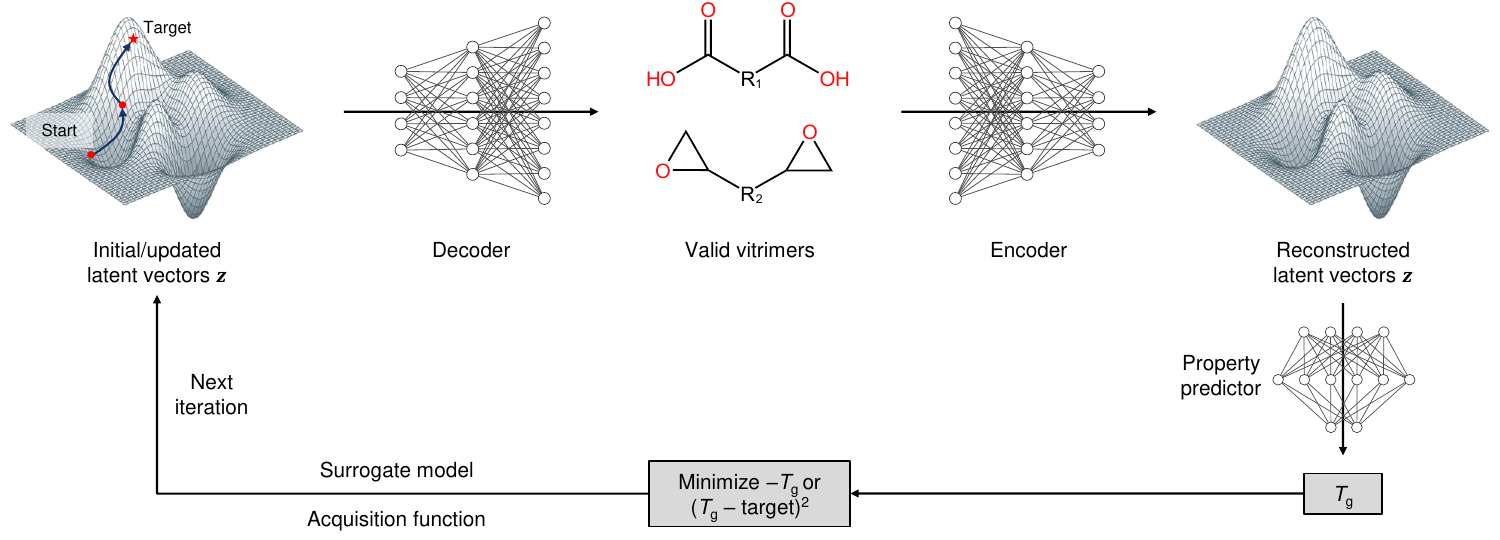}
\caption{\textbf{Schematic workflow of Bayesian optimization.}}
\label{boflow}
\end{figure}

\begin{figure}[!ht]
\centering
\includegraphics[width=\linewidth]{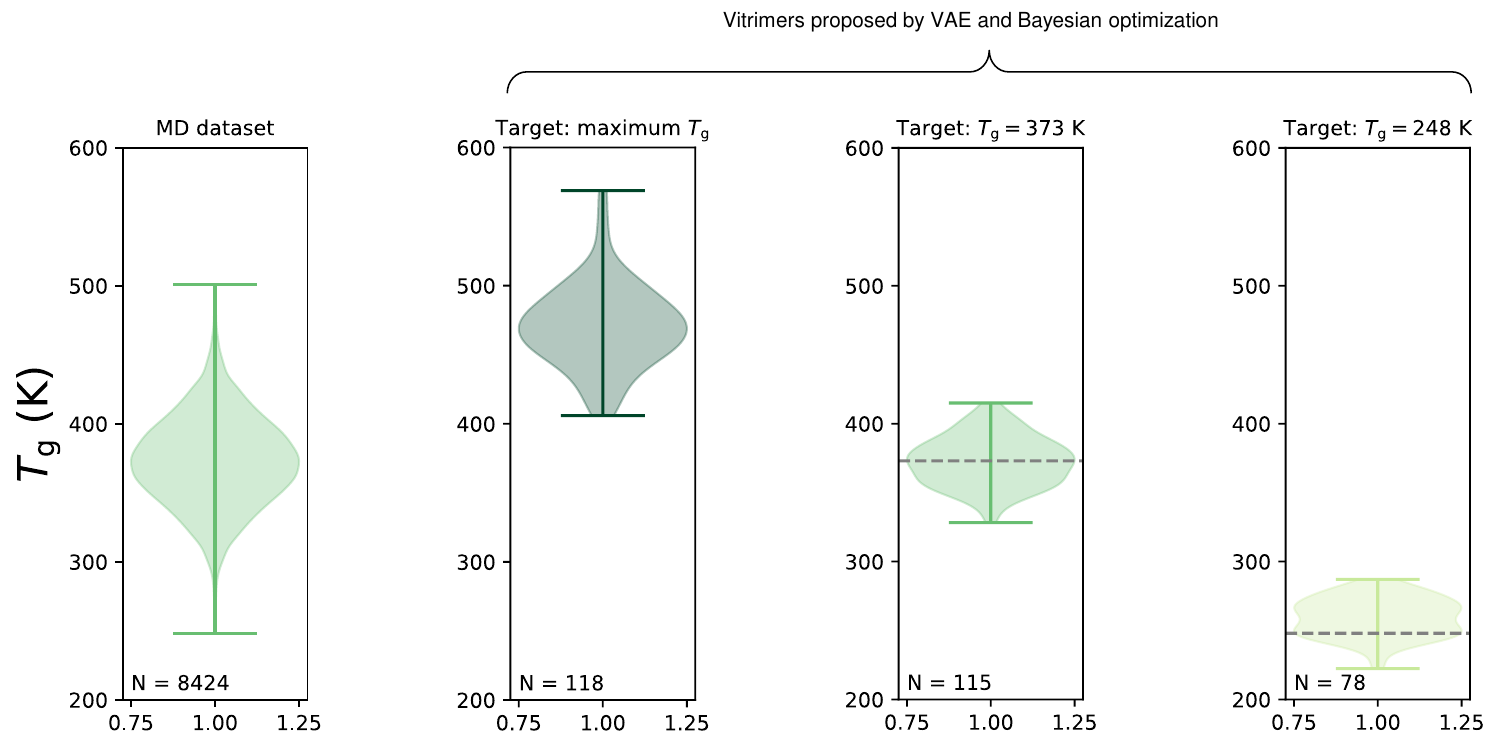}
\caption{\textbf{Distributions of $T_\mathrm{g}$ of the dataset and the vitrimers proposed by the VAE and Bayesian optimization.} All presented $T_\mathrm{g}$ values of are validated by MD simulations and GP calibration.}
\label{violin}
\end{figure}

\begin{figure}[!ht]
\centering
\includegraphics[width=\linewidth]{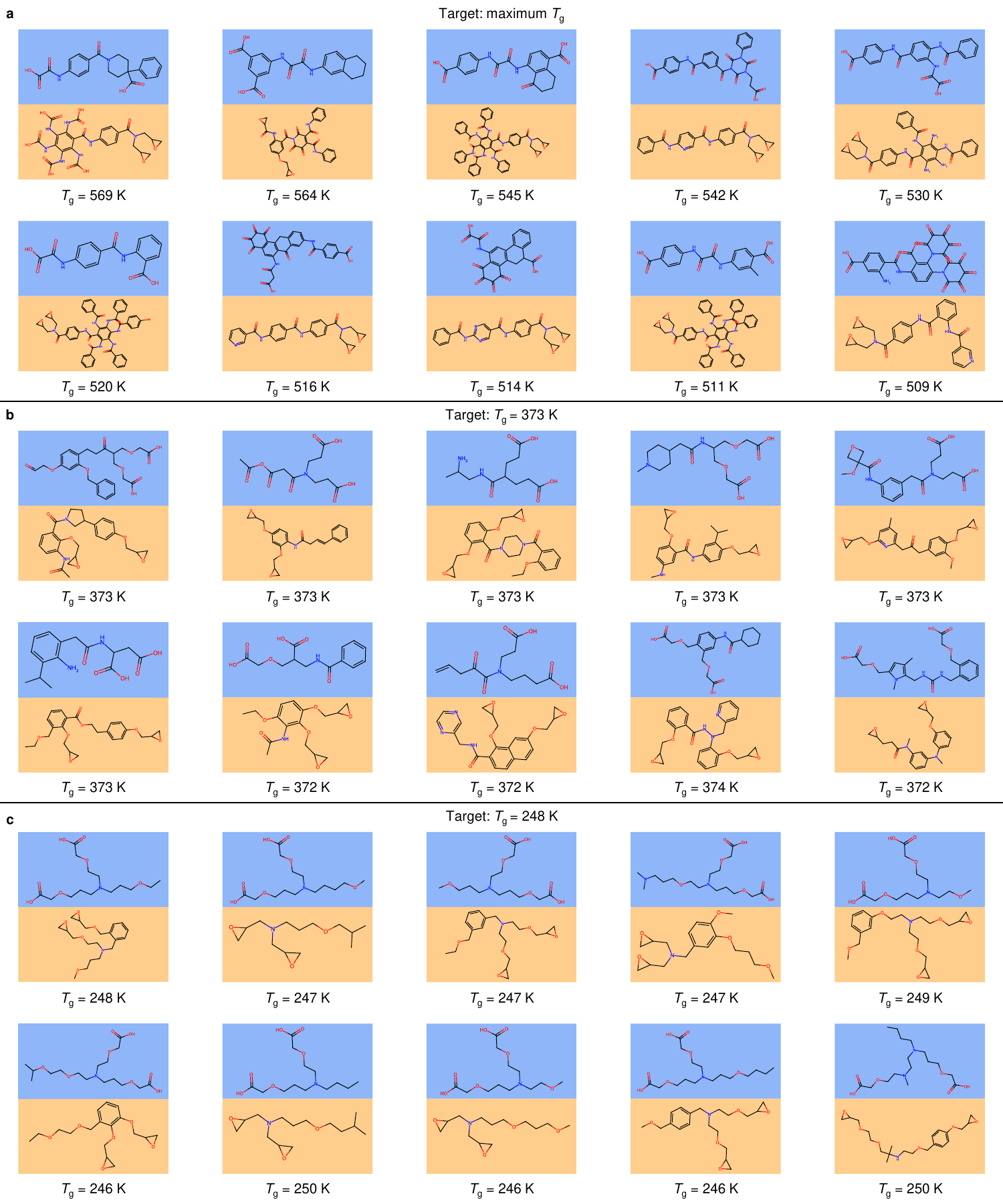}
\caption{\textbf{Examples of novel vitrimers designed with different target $T_\mathrm{g}$ from Bayesian optimization.} \textbf{(a)} Maximum $T_\mathrm{g}$. \textbf{(b)} Target $T_\mathrm{g} = 373~\mathrm{K}$. \textbf{(c)} Target $T_\mathrm{g} = 248~\mathrm{K}$. All presented $T_\mathrm{g}$ values of proposed vitrimers are validated by MD simulations and GP calibration.}
\label{morebo}
\end{figure}

\begin{figure}[!ht]
\centering
\includegraphics[width=\linewidth]{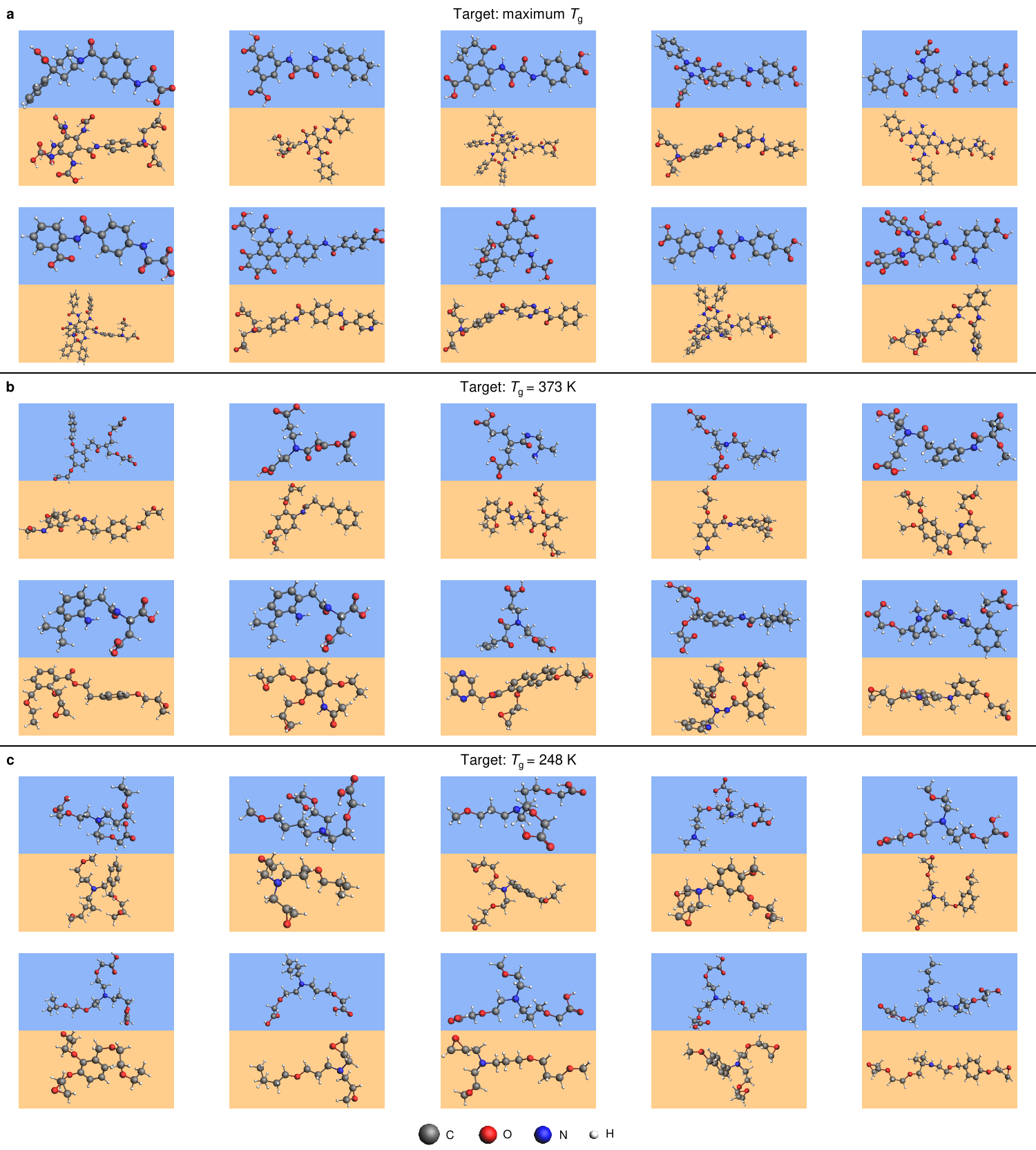}
\caption{\textbf{Minimized structures of vitrimer components presented in Supplementary Figure \ref{morebo}.} The molecules are minimized by ReaxFF using the CHON2017\_weak\_bb force field.}
\label{minimized}
\end{figure}

\begin{figure}[!ht]
\centering
\includegraphics[width=\linewidth]{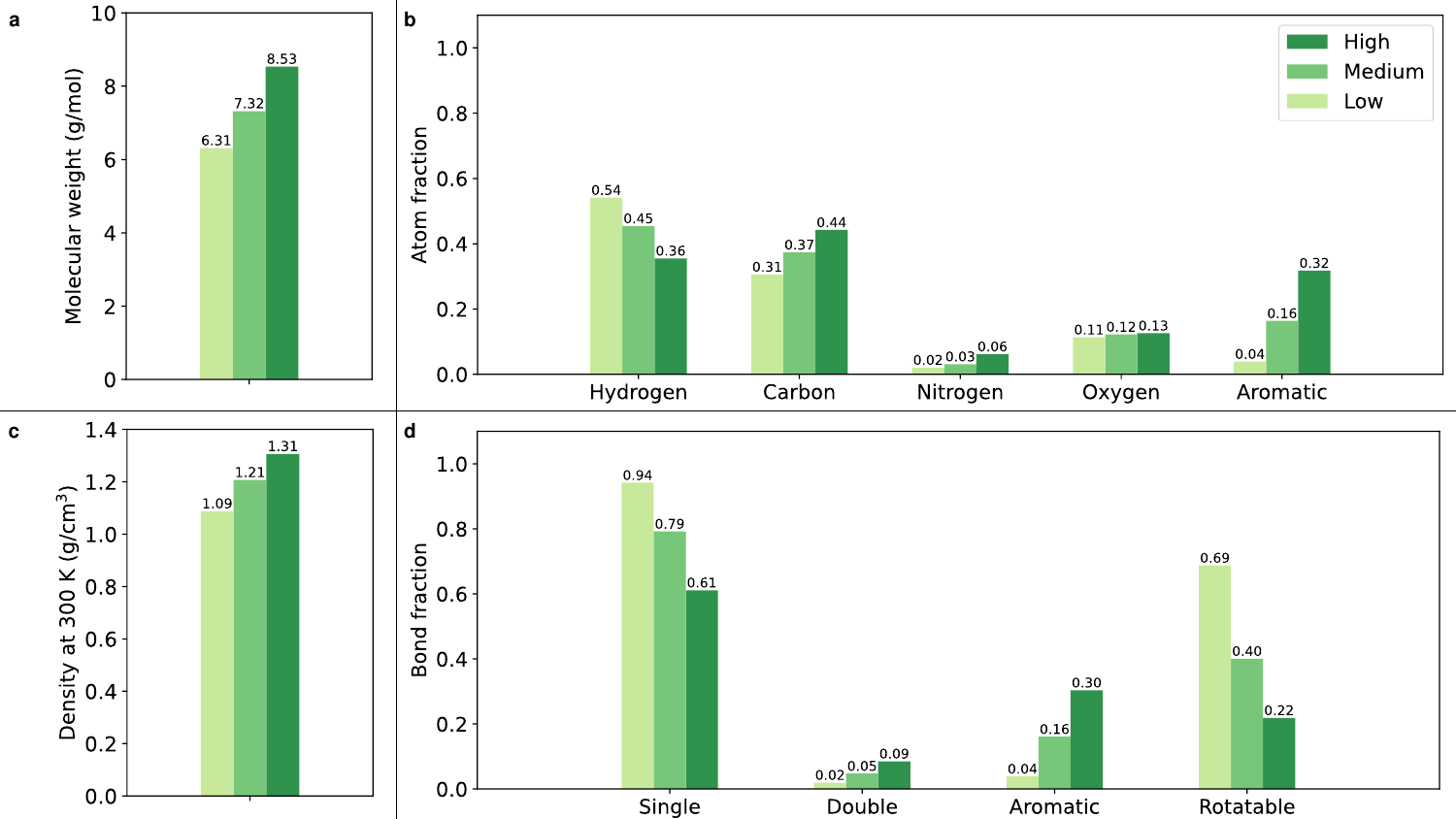}
\caption{\textbf{Relevant molecular descriptors of the vitrimers presented in Supplementary Figure \ref{morebo}.} \textbf{(a)} Molecular weight. \textbf{(b)} Atom fractions. \textbf{(c)} Density at 300 K. \textbf{(d)} Bond fractions. \textbf{}All values are averages of ten vitrimers per target.}
\label{descriptor}
\end{figure}

\begin{figure}[!ht]
\centering
\includegraphics[width=\linewidth]{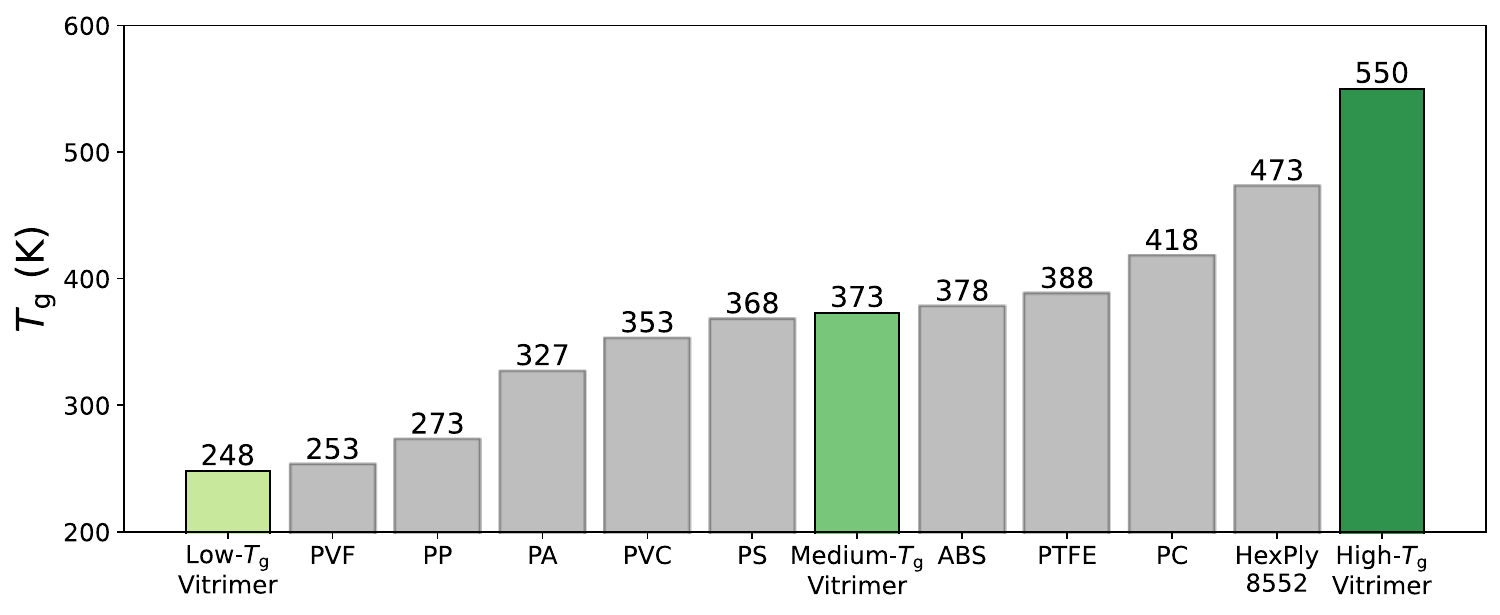}
\caption{\textbf{$T_\mathrm{g}$ of commonly used polymers and our proposed vitrimers by inverse design.}}
\label{tgbarchart}
\end{figure}

\section{Computational efficiency} \label{supp_efficiency}

\begin{table}[t]
    \centering
    \begin{tabular}{c c c c}
        \Xhline{2\arrayrulewidth}
        \textbf{Tasks} & \textbf{Software} & \textbf{Hardware} & \textbf{Runtime (h)} \\
        \hline
        $T_\mathrm{g}$ calculation of one vitrimer by MD & LAMMPS & CPU & 310\\
        Training of the VAE with $\mathcal{D}$ & PyTorch & GPU & 130\\
        Joint training of the VAE and property predictor $\mathcal{F}_\omega$ with $\mathcal{D}_\mathrm{prop}$ & PyTorch & GPU & 5\\
        $T_\mathrm{g}$ prediction of 1,000 vitrimers by trained $\mathcal{F}_\omega$ & PyTorch & GPU & 0.01\\
        Proposing $\sim$ 1,300 candidate vitrimers by Bayesian optimization & PyTorch & GPU & 1\\
        \Xhline{2\arrayrulewidth}
    \end{tabular}
    \caption{\textbf{The computational runtime, used software and hardware for different tasks in this work.} The presented runtime is roughly estimated with one core of a 2.1 GHz Intel Xeon Gold 6230 CPU and a NVIDIA GeForce RTX 2080 Ti GPU.}
    \label{runtime}
\end{table}

The computational runtime, used software and hardware for different tasks are listed in Supplementary Table \ref{runtime} to demonstrate the efficiency of the proposed method in this work. The trained property predictor serves as a shortcut for costly MD simulations to estimate $T_\mathrm{g}$ of vitrimers, which also allows for efficient discovery of novel vitrimers with target $T_\mathrm{g}$ by Bayesian optimization.

\section{Experimental synthesis and characterization} \label{supp_experiment}

\begin{figure}[t]
\centering
\includegraphics[width=.5\linewidth]{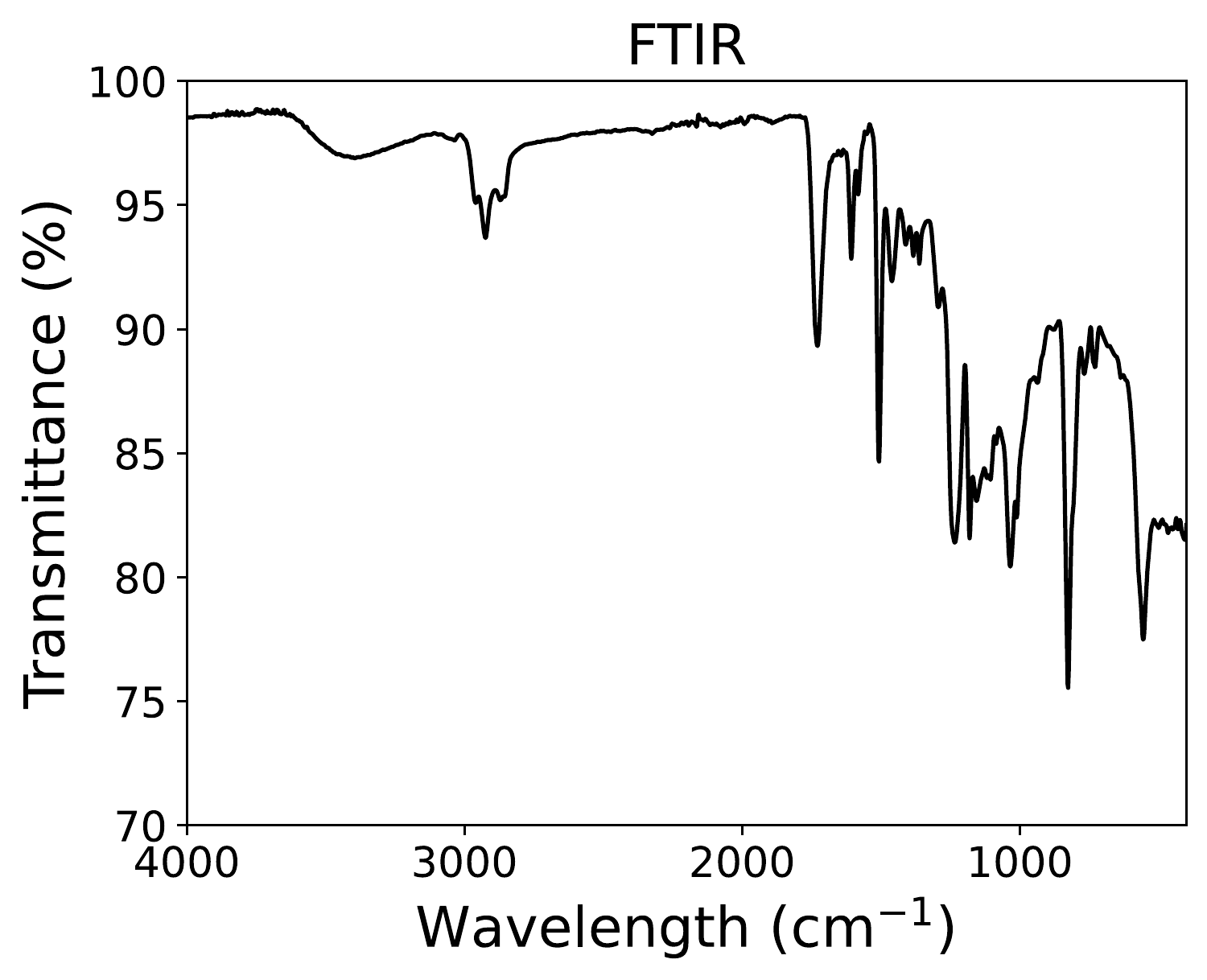}
\caption{\textbf{Fourier-transform infrared (FTIR) spectroscopy of the synthesized vitrimer.}}
\label{FTIR}
\end{figure}

The synthesis of the proposed vitrimer (Figure \ref{synthesis} in the manuscript) is performed by a two-step reaction. The first step involves mixing of succinic anhydride (1 mol) with glycerol (0.5 mol) at 130\textdegree C for 1 hour to obtain a clear solution mixture of polymer with bifunctional carboxylic acid groups. In the second step, equivalent stoichiometric amount of DGEBA is added to the mixture in presence of catalyst triazabicyclodecene (TBD), which is further mixed at the same condition for 30 minutes to obtain a homogeneous slurry. The slurry is then transferred into a preheated Teflon taped mold at 145\textdegree C and covered with top cover. The mold is placed into the heat press to continue the crosslinking at 145\textdegree C for 6 hours followed by post curing for 2 hours at 180\textdegree C.

The crosslinked vitrimer specimen is verified using Fourier-transform infrared (FTIR) spectroscopy, as shown in Supplementary Figure \ref{FTIR}. Differential scanning calorimetry (DSC) is carried out by TA Instruments Q2000. Thermomechanical analysis (TMA) is performed at 0.1N applied force with a heating rate of 5\textdegree C per minute using Perkin Elmer TMA7. 

\clearpage

\bibliography{sample}

\makeatletter\@input{xx.tex}\makeatother